\documentclass[letterpaper]{mn2e}
\usepackage{savesym}
\savesymbol{bibhang}
\usepackage{natbib}
\restoresymbol{TXF}{bibhang}
\usepackage{graphicx}
\usepackage{epsfig}
\usepackage{times}
\usepackage{amsfonts}
\usepackage{subfigure}
\usepackage{pdflscape}
\usepackage{longtable}
\usepackage{chngcntr}
\usepackage[asymmetric,tmargin=0.76in,bmargin=0.82in,lmargin=0.61in,rmargin=0.93in]{geometry}

\newcommand {\apgt} {\ {\raise-.5ex\hbox{$\buildrel>\over\sim$}}\ }
\newcommand {\aple} {\ {\raise-.5ex\hbox{$\buildrel<\over\sim$}}\ }

\newcommand {\aj} {AJ}
\newcommand {\apj} {ApJ}
\newcommand {\apjl} {ApJ}
\newcommand {\apjs} {ApJS}
\newcommand {\aap} {A\&A}
\newcommand {\aaps} {A\&AS}
\newcommand {\mnras} {MNRAS}

\newcommand {\araa} {ARA\&A}

\title[Barred S0 Galaxies in the Coma Cluster]
{Barred S0 Galaxies in the Coma Cluster}

\author[George Lansbury et al.]
{George B. Lansbury$^{1}$\thanks{Email: g.b.lansbury@durham.ac.uk}, John
  R. Lucey$^{1}$, Russell J. Smith$^{1}$
\\
$^1$Department of Physics, University of Durham, South Road, Durham DH1 3LE\\
}

\begin{document}
{\rm {\bf {\rm }}}
\maketitle


\begin{abstract}
This study uses $r$-band images from the Eighth Data Release of the Sloan Digital Sky Survey (SDSS DR8) to study bars in lenticular (S0) galaxies in one of the nearest rich cluster environments, the Coma cluster. 
We develop techniques for bar detection, and assess their success when applied to SDSS image data. To detect and characterise bars we perform 2D bulge+disk+bar light decompositions of galaxy images with GALFIT. Using a sample of artificial galaxy images we determine the faintest magnitude at which bars can be successfully measured at the depth and resolution of SDSS. We perform detailed decompositions of $83$ S0 galaxies in Coma, $64$ from a central sample, and $19$ from a cluster outskirts sample. For the central sample, the S0 bar fraction is $72^{+5}_{-6} \%$. This value is significantly higher than that obtained using an ellipse fitting method for bar detection, $48^{+6}_{-6}\%$. At a fixed luminosity, barred S0s are redder in $(g-r)$ colour than unbarred S0s by $0.02$~mag. The frequency and strength of bars increase towards fainter luminosities. Neither central metallicity nor stellar age distributions differ significantly between barred and unbarred S0s. 
There is an increase in the bar fraction towards the cluster core, but this is at a low significance level. Bars have at most a weak correlation with cluster-centric radius. 
\end{abstract}


\section{Introduction}
\label{Introduction}

Stellar bars are effective drivers of secular evolution in disk galaxies. The rotation of bars couples to the motion of galactic material, creating characteristic rings and dense central regions \citep{Sellwood93,Kormendy04}, and redistributing angular momentum between the disk and dark matter halo \citep{Weinberg85,Debattista98,Debattista00,Bournaud02,Athanassoula03,MartinezValpuesta06,Berentzen07,Athanassoula13}. As a result of this redistribution bars can grow in strength, becoming increasingly efficient at funnelling gas towards central regions where starbursts may occur \citep{Hawarden86,Martinet97,Regan04,Jogee05,Ellison11} and the formation of bulges or pseudo-bulges may be augmented \citep{Kormendy04,Athanassoula05,Jogee05,Gadotti11}.

In terms of bar-driven radial gas inflows and their influence on star formation, chemical enrichment, and bulge formation, spectroscopic studies of galaxy centres are of key importance. Such studies have shown that barred spiral galaxies have enhanced star formation rates relative to their unbarred counterparts 
\citep{Ho97,Jogee05,Ellison11}. With regards to central metallicities, conflicting results have been obtained. 
For instance, \citet{Coelho11} find similar stellar metallicities for barred and unbarred galaxies, whilst \citep{Perez11} report higher metallicities in barred galaxies. Thus, the question of whether bars influence chemical enrichment in galaxy centres is a matter of debate. 
While studies of central stellar ages are few and mostly limited by small samples \citep[e.g.,][]{Perez11}, \citet{Coelho11} present stellar population analyses for a statistical sample which includes all disk galaxy types. They measure, at a significance level of $\sim$$4\sigma$, that barred galaxies have on average younger central stellar populations than unbarred galaxies; evidence for bars playing an important role in the building of bulges. 

The question of how the formation and evolution of bars are affected by environment is also the subject of debate. For example, some numerical simulations show that fly-by tidal interactions of the type found in dense clusters should be able to induce bars for specific orbital configurations \citep{Romano08,Aguerri09b}, while other studies suggest that fast, frequent and weak galaxy encounters can dynamically heat disks, making them less prone to the disk instabilities which lead to bar formation \citep{Aguerri09b,Kormendy12}. 
Observations of bars in the extremely dense environments of cluster cores, and comparison with lower density environments, can therefore provide valuable information in helping understand the relative contributions of internal and external processes to the dynamical evolution of disk galaxies.

Most observational studies find weak to no variation of bars across environments of varying density \citep{vandenBergh02,Barazza09a,Marinova09,Aguerri09a,Mendez-Abreu10,Martinez11,Mendez-Abreu12}. One study to report a significant bar-environment correlation is that of \citet{Skibba12}, whose 
result that barred galaxies are more likely to be found in denser environments is significant at the $>6\sigma$ level. 
At least half of their measured correlation is contributed to by colour-environment and stellar mass-environment dependences, as opposed to the direct influence of environmental processes.
Some studies suggest a radial increase in the bar fraction towards the dense cores of clusters \citep{Andersen96,Barazza09a,Barazza09b,Marinova12}, but such results are limited by small samples. 

Recent studies have investigated correlations between bars and many other galaxy properties; luminosity, colour, effective radius, central velocity dispersion, stellar mass, bulge-to-total ratio, Hubble type and redshift \citep[recent examples include][]{Weinzirl09b,Cameron10,Mendez-Abreu10,Barway11,Masters11,Masters12,Laurikainen13}.
As a relevant example, \citet{Laurikainen09} find that lenticulars, i.e. S0 galaxies, have a mean bar fraction less than that of spiral galaxies, 
and \citet{Buta08,Buta10} measure considerably weaker bars in lenticulars than in spirals.


In this paper, we investigate bars in lenticulars in the Coma cluster. By studying galaxies that lie at different cluster radii, a wide range of environments can be probed \citep{Lucey91}. Lenticulars are the dominant morphological galaxy type in the cores of nearby rich clusters such as Coma \citep{Dressler80}. This relatively large abundance of lenticulars, and a high central galaxy density make Coma an excellent laboratory for studying the environmental dependence of bars. 

In order to develop and test techniques for bar detection and characterisation, 
we perform detailed structural analyses of $64$ S0 galaxies within the central $1.5^{\circ}$ ($2.5$~Mpc) radius region of Coma. We also analyse a control sample of $19$ S0 galaxies that are associated with Coma but lie $\sim$$10$~Mpc from the cluster core.
We refer to these two samples as the `central' and `outskirts' samples, respectively. Optical image data is from the Eighth Data Release of the Sloan Digital Sky Survey \citep[SDSS DR8,][]{Aihara11}. We use the two-dimensional (2D) profile fitting algorithm GALFIT \citep{Peng02} to decompose galaxy images into bulge, disk and bar components. First, we verify our method by performing bulge+disk+bar decompositions for a sample of artificial galaxies. We use this artificial galaxy fitting to determine a magnitude limit for the successful recovery of parameters, and investigate the residual flux fraction ($\rm RFF$) goodness-of-fit parameter \citep{Hoyos11} as a quantitative bar detection parameter.
Second, we perform decompositions of our S0 samples. We subsequently present results for the dependence of the bar fraction ($f_{\mathrm{bar}}$), bar probability ($p_{\mathrm{bar}}$), and bar strength ($\Phi_{\mathrm{bar}}$) on environment and on galaxy properties. To study the influence of bars on central stellar metallicities and ages, we compare our bar analysis with the spectroscopic measurements of \citet{Smith12}. We interpret the results in the context of bar-driven gas inflows.

The paper is organised as follows. We describe our methodology for bar detection and characterisation, including the image decomposition procedure (Section \ref{fitting_routine}) and bar detection criteria (Section \ref{bar_criteria}). Our galaxy sample selection and data set are detailed in Section \ref{sample}. Results from the analysis of our sample are presented (Section \ref{Results}), and their implications are discussed in the context of other studies (Section \ref{Discussion}). 
We present our main conclusions in Section \ref{Conclusions}. We adopt a physical distance to the Coma cluster of $100$~Mpc and a scale of $0.483$~kpc arcsec$^{-1}$ \citep[c.f.,][]{Carter08}.

\section{Bar Detection}
\label{method}

\subsection{Introduction}
\label{method_intro}

Early attempts to measure bars in disk galaxies used visual examination and a subsequent classification of galaxies as either strongly-barred, weakly-barred or unbarred \citep[e.g.,][]{deVaucouleurs91,Eskridge00}.\label{weakstrong} 
A number of more sophisticated methods have since been developed which attempt to define a continuous, measurable parameter to represent bar `strength'.  
One such technique, developed by \citet{Martin95}, involves the fitting of ellipses to galaxy isophotes. \label{IRAF_ellipse_method} If the following criteria are met, galaxies are considered barred: (1) outwards from the galaxy centre ellipticity ($e$) rises steadily to a global maximum greater than $0.25$ and the PA stays within $\pm10^{\circ}$, (2) after the global maximum, $e$ drops by a minimum of $0.1$ and the PA changes by more than $10^{\circ}$ \citep[e.g.,][]{Barazza09a,Marinova10}. 
In this method, the maximum ellipticity $e_{\mathrm{bar}}$ can be used as a basic parameter for bar strength.

More recently bar measurement has been achieved using bulge+disk+bar decomposition \citep{Prieto97,Prieto01,Aguerri05,Weinzirl08,Weinzirl09a,Gadotti11}, which involves the 2D modelling of galaxy light distributions with bulge, disk and bar components. Examples of code designed to fit such components include GALFIT \citep{Peng02} and BUDDA \citep{deSouza04}. The method yields many structural parameters, including bar ellipticity $e_{\mathrm{bar}}$ and bar light fraction $Bar/T$, each of which is a partial measure of bar strength. As such, $\Phi_{\mathrm{bar}}=e_{\mathrm{bar}}\times Bar/T$ can be used as a combined, quantitative measure of bar strength \citep{Weinzirl09b}. In this work, we adopt this bulge+disk+bar decomposition method as a means of detecting and characterising bars.


\subsection{GALFIT Decomposition Procedure}
\label{fitting_routine}
We use the 2D surface fitting routine GALFIT, developed by \citet{Peng02,Peng10}, to perform bulge+disk+bar decomposition. GALFIT is a non-linear, least-squares fitting algorithm that uses the Levenberg-Marquardt algorithm to find $\chi^{2}$ minima, given initial parameter guesses. 
In our implementation, we provide GALFIT with a pre-calculated sky background component, 
fitting region specifications, a PSF image for convolution with the model, an external object or bad pixel mask file, and a sigma image (noise map). 
We use pre-calculated sigma images, having found that they produce slightly more reliable results when fitting artificial galaxy images than those automatically generated by GALFIT. We employ an iterative procedure which follows that of \citet{Weinzirl09b}, whereby structural components are successively added to the GALFIT model. The stages of the fitting procedure are as follows:

\begin{enumerate}
\item S\'ersic fit: The galaxy image is fit using a single S\'ersic component with a radial surface-brightness profile of:
\begin{equation}
\Sigma(r) = \Sigma_{e} \exp \left[-\kappa\left(\left(\frac{r}{r_{e}}\right)^{1/n}-1\right)\right],
\end{equation}
where $\Sigma_{e}$ is the surface brightness at the effective radius $r_{e}$ (i.e. the radius enclosing half of the total flux), $n$ is the S\'ersic index, and $\kappa$ is a dependent variable coupled to $n$ such that half of the total flux is enclosed within $r_{e}$ \citep[see][]{Graham05}.
Initial parameter estimates do not need to be precise at this stage as GALFIT easily converges on a solution.

\item bulge+disk fit: To the single S\'ersic (bulge) component we add an exponential disk component with a radial profile of:
\begin{equation}
\Sigma(r) = \Sigma_{0} \exp \left(-\frac{r}{r_{s}}\right),
\end{equation}
where $\Sigma_{0}$ is the central surface brightness and $r_{s}$ is the scale length of the disk, related to the effective radius through the relationship $r_{e}=1.678r_{s}$. A bulge+disk fit is performed, with the disk axial ratio [$(b/a)_{\mathrm{disk}}$] and position angle ($\mathrm{PA}_{\mathrm{disk}}$) set to values measured using the IRAF task \verb|ellipse|. 

\item bulge+disk+bar fit: To the bulge+disk model we add a low index ($n=0.5$) S\'ersic component representing a bar. Initial guesses for $(b/a)_{\mathrm{bar}}$ and $\mathrm{PA}_{\mathrm{bar}}$ are deduced using \verb|ellipse|. After this bulge+disk+bar fit is performed, the $(b/a)_{\mathrm{disk}}$ and $\mathrm{PA}_{\mathrm{disk}}$ parameters are freed such that GALFIT may reach a stable solution.
\end{enumerate}


\subsection{Determination of a Magnitude Limit for Sample Selection}
\label{magnitude_limit}

To determine a suitable magnitude limit for sample selection, we investigated the optimum signal-to-noise ratio ($S/N$) at which bar parameters can be reliably measured using the GALFIT decomposition procedure outlined in Section \ref{fitting_routine}. First, we fitted noise-added model SB0 galaxy images designed to mimic SDSS DR8 $r$-band data in terms of resolution and $S/N$ range. In general, GALFIT bar parameters failed to be recovered below $S/N\sim100$.
Second, we applied the decomposition procedure to $50$ model SB0 galaxies with $S/N=100$; $25$ of which have $Bar/T=10\%$ and $25$ of which have $Bar/T=20\%$. 
To determine how well the $Bar/T$ parameter is recovered at $S/N=100$, 
we calculated the ratio of the best-fit value of $Bar/T$ to the original model value (Fit/Model) for each individual galaxy, and subsequently the overall standard deviation of Fit/Model ($\sigma_{\mathrm{Fit/Model}}$). 
For $Bar/T=20\%$ an acceptable scatter of $\sigma_{\mathrm{Fit/Model}}=0.15$ was measured, but for $Bar/T=10\%$ a high scatter of $\sigma_{\mathrm{Fit/Model}}=0.37$ supports that $S/N=100$ is an appropriate signal-to-noise limit for sample selection. 
This corresponds to a magnitude limit of $r_{\mathrm{petro}}=16.7$ for SDSS $r$-band data. 


\subsection{Bar Detection Criteria}
\label{bar_criteria}

In our analysis we require that the following three criteria are satisfied for a galaxy to be classified as barred:

\begin{enumerate}
\item \label{visual_residual} A bar must be visually identified in the S\'ersic and bulge+disk model subtracted image residuals that is removed when a bar is added to the model. Clearly, identifying such signatures by eye is a subjective method.
To address this issue, we have generated a large set of artificial galaxy images with and without bars, and analysed these following the procedure outlined in Section \ref{fitting_routine}. The best-fit model subtracted residual images provide a reference set for visual comparison to real galaxy residual images.  
Some example residual images for single S\'ersic fits are displayed in Fig. \ref{residual_comb_diagram}, where galaxies have been arranged according to their $e_{\rm bar}$ and $Bar/T$ values. The diagram only shows results for one set of bulge and disk parameters and is not intended to cover the complete range of possible residual signatures, but to give an idea of typical bar signatures and how these can be expected to vary with bar strength. In an additional simplification $\mathrm{PA}_{\mathrm{bulge}}$ and $\mathrm{PA}_{\mathrm{disk}}$ have been set equal, which is often not a good approximation for barred galaxies. The bar produces a distinctive pattern in the residual images. 
\begin{figure*}
\begin{minipage}{\textwidth}
\centering
\includegraphics[width=0.9\textwidth]{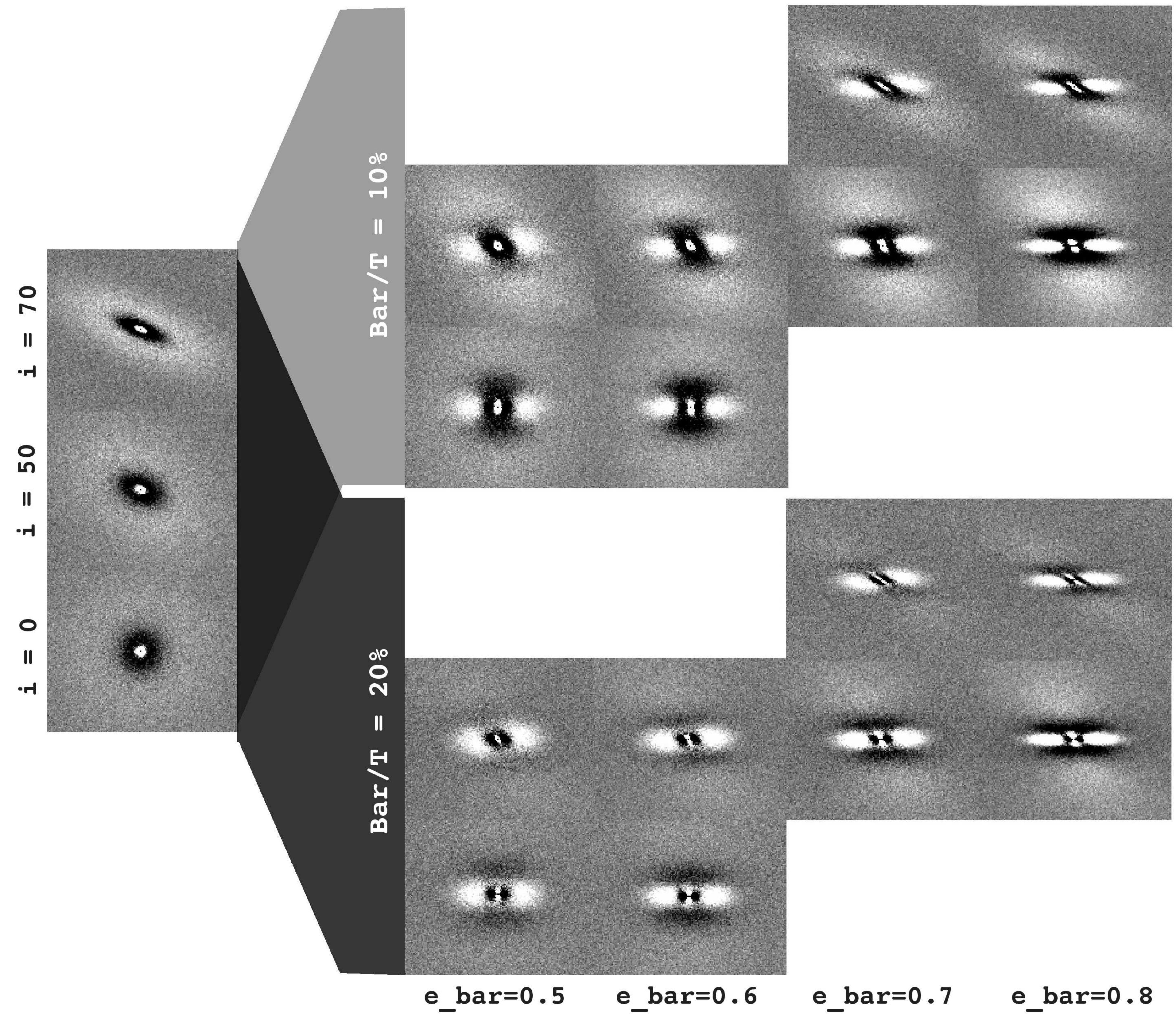}
\end{minipage}
\begin{minipage}{\textwidth}
\centering
\caption{Diagram showing the effect on single S\'ersic fit image residuals of adding different strength bar components (right) to a simple bulge+disk model (far left). For these model galaxies, $S/N\sim200$. The images are grouped by bar light fraction with $Bar/T=10\%$ above and $20\%$ below, increasing with $e_{\mathrm{bar}}$ from left to right and $i$ from bottom to top. Bars are horizontally oriented to aid the comparison of residual signatures. Columns do not correspond to different inclinations of the same galaxy.}
\label{residual_comb_diagram}
\end{minipage}
\end{figure*}

\item \label{sensible_parameters} Best-fit parameters must take on sensible values for the bulge+disk+bar fit to be accepted, i.e. they must lie within typical parameter ranges and not converge to unreasonably high or low values. For example, a bulge+disk+bar fit will only be accepted if $n_{\mathrm{bar}}\sim0.5$ and $n_{\mathrm{bulge}}\apgt1$, the lower limit corresponding to a pseudo-bulge \citep{Gadotti09}.

\item For the third criterion, we define a bar detection parameter ($\Delta\mathrm{RFF}$) that quantifies the change in image residuals between the bulge+disk and bulge+disk+bar fitting stages, thereby increasing objectivity.

The residual flux fraction ($\mathrm{RFF}$) measures the fraction of the image residuals which cannot be accounted for by noise, and is defined by \citet{Hoyos11} as:
\begin{equation}
\mathrm{RFF} = \frac{\Sigma_{i}|\mathrm{Res}_{i}|-0.8\times\sigma_{\mathrm{image}}}{\mathrm{FLUX\_ISO}},
\label{RFF_equation}
\end{equation}
where $|\mathrm{Res}_{i}|$ is the modulus of the remaining pixel value after subtraction of the best-fit model from the original image, the summation of which is over all pixels within the galaxy iso-area. 
$\sigma_{\mathrm{image}}$ is the image variance, and FLUX\_ISO is the total flux of the iso-area. For the iso-area we use the area of the moments ellipse as defined by SExtractor \citep{Bertin96}. A very small or negative $\mathrm{RFF}$ is interpreted as over-fitting. 
 \citet{Hoyos11} find that an $\mathrm{RFF}$ of greater than $11\%$ justifies the addition of further model components. Our bulge+disk and bulge+disk+bar results very rarely exceed this fraction, and there is always at least a small decrease in $\mathrm{RFF}$ as model components are added. As such, the change in $\mathrm{RFF}$ between the bulge+disk and bulge+disk+bar image residuals ($\Delta \mathrm{RFF}$) was instead investigated as a parameter for indicating whether a galaxy is likely to be barred or unbarred.

To gauge typical $\Delta \mathrm{RFF}$ values for barred and unbarred lenticulars (SB0s and S0s, respectively), we fitted $2000$ artificial galaxy images ($1000$ of each type), designed to mimic our SDSS sample in terms of $S/N$ range, resolution, and the physical properties (light fractions, effective radii, axis ratios and S\'ersic indices) of the morphological components. The resulting $\Delta \mathrm{RFF}$ distributions are shown in Fig. \ref{delta_RFF_histograms}{\it a}. For comparison, we show the $\Delta \mathrm{RFF}$ distributions for real SDSS images of Coma cluster galaxies with visually identified bars (Fig. \ref{delta_RFF_histograms}{\it b}). Similar ranges in $\Delta \mathrm{RFF}$ are covered. We adopt a bar detection threshold of $\Delta \mathrm{RFF} = 0.5\%$, since $90\%$ of the model SB0s lie above, and $93\%$ of the model S0s lie below this value.
Thus, our third bar detection criterion is that $\Delta \mathrm{RFF}$ must be greater than $0.5\%$ for a galaxy to be classified as barred.
\begin{figure}
\centering
\includegraphics[width=0.5\textwidth]{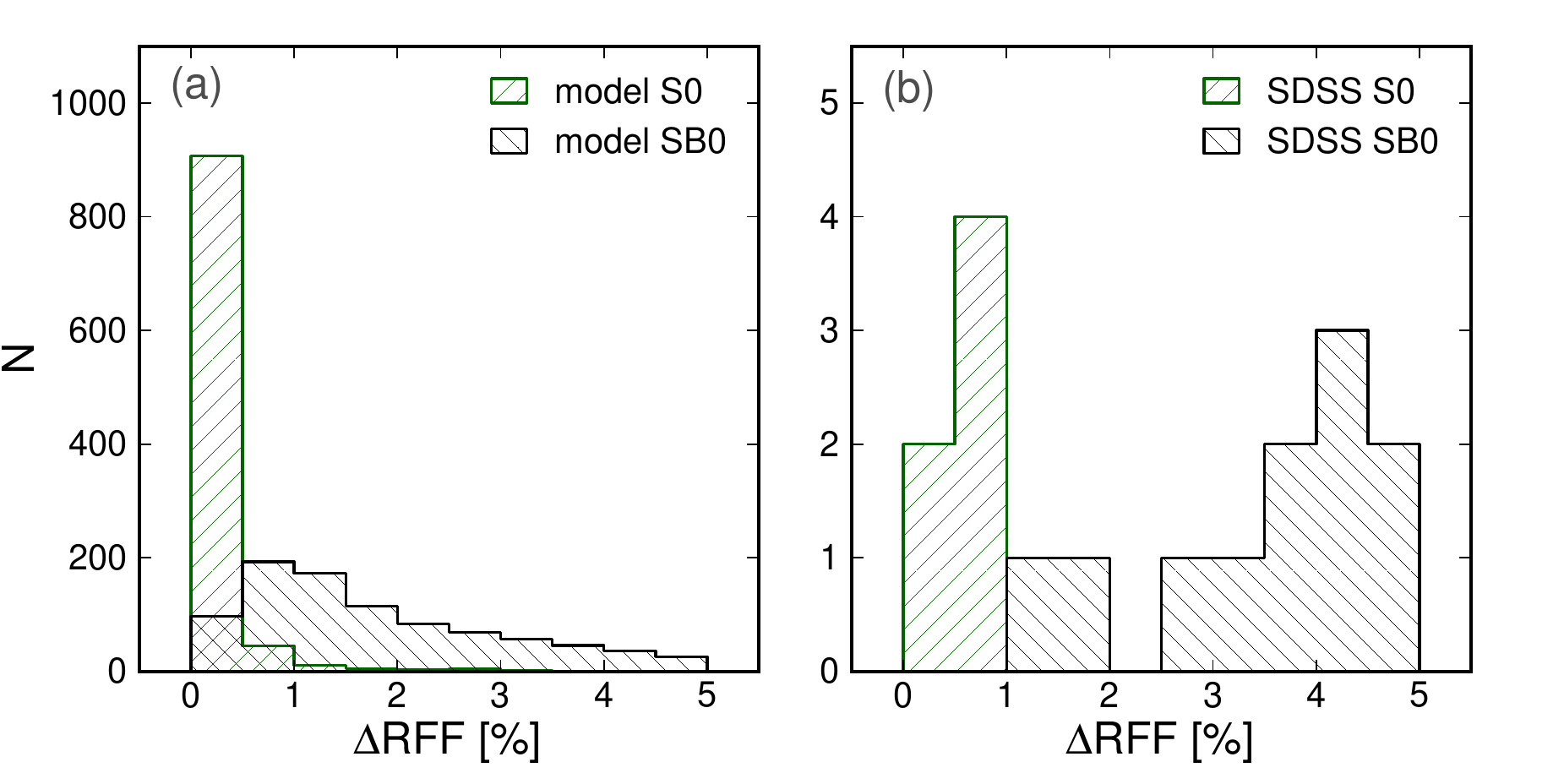}
\caption{Distributions of the bar detection parameter $\Delta \mathrm{RFF}$ for barred and unbarred lenticulars (SB0s and S0s, respectively), using both: (a) model galaxy images, and (b) SDSS galaxy images with visually identified bars. Since $90\%$ of the model SB0s have $\Delta \mathrm{RFF}$ greater than $0.5\%$, we use this as a threshold for bar detection (see Section \ref{bar_criteria}).}
\label{delta_RFF_histograms}
\end{figure}
\end{enumerate}

\subsection{Application of Bar Detection Method to Model Galaxies}
\label{model_detection}

Fitting artificial galaxies is an important step in assessing our bar detection method. We applied the decomposition procedure in Section \ref{fitting_routine} and the bar detection criteria in Section \ref{bar_criteria} to $2000$ model galaxy images, $1000$ S0s and $1000$ SB0s, designed to mimic our real galaxy sample in terms of $S/N$ and morphological properties (for a similar analysis, see Section 4 of \citealt{Aguerri09a}). Our method efficiently identifies bars, with $85.7\%$ of the model SB0s correctly identified as barred, and only $14.3\%$ incorrectly identified as unbarred. Of the model S0s, $99.3\%$ were correctly identified as unbarred, and $0.7\%$ were incorrectly identified as barred. As such, the bar fractions given in Section \ref{Results} may be lower limits.

\section{Data set and Galaxy Sample}
\label{sample}

We study exclusively lenticular (S0) galaxies in the Coma cluster using $r$-band images from the SDSS DR8. 
As noted in Section \ref{Introduction}, the bar fraction and bar properties have been shown to vary significantly with Hubble type. The abundance of S0s in Coma 
thus allows a statistically robust sample of one specific disk-galaxy type, removing selection effects caused by this variation. Additionally, the lack of spiral features in S0s makes them well suited to our bar detection method (see \citealt{Aguerri12} for a review of the photometric components of S0s).

Comparisons of bars between different density environments often consider results from separate studies. This may limit the conclusions from any measured bar-environment correlation as bar definitions tend to be based on measurable parameters associated with the specific method used, and the number of detectable bars increases significantly with $S/N$ \citep[e.g.,][]{Menendez-Delmestre07}. SDSS DR8 data is therefore particularly useful, covering the entire Coma cluster field and allowing a self-consistent study of bars spanning a wide range of environments.


\subsection{Sample Selection}
\label{sample selection}

The main `central' sample of Coma cluster S0s was selected as follows.
The cluster centre was taken as the mid-point between NGC 4874 and NGC 4889, $(\mathrm{RA}~=~194.9663^{\circ}, \mathrm{Dec}~=~27.9681^{\circ})$. SDSS galaxies within a $1.5^{\circ}$ ($2.5$~Mpc) radius of the centre were selected. While the virial radius of Coma is $\sim$$2.9$~Mpc \citep{Lokas03}, within this radius gradients in the properties of luminous galaxies are observed \citep{Smith12}. 
Cluster membership was determined using SDSS DR8 spectroscopic redshifts and 
the caustic pattern calculated by \citet{Rines03}. 
A colour cut was made of $g-r>0.6$, corresponding to the red sequence. An initial magnitude cut was applied at $r_{\mathrm{petro}}<16.7$, an upper limit for the successful measurement of bars determined through artificial galaxy fitting (Section \ref{magnitude_limit}). This selection resulted in a sample of $395$ galaxies.

An inclination cut was applied as bulge+disk+bar decomposition fails to give reasonable results for highly inclined galaxies. Galaxies with disk ellipticity ($e_{\mathrm{disk}}$) $>0.5$, which is equivalent to inclination $>60^{\circ}$, were identified through isophotal analysis with the IRAF task \verb|ellipse| and subsequently removed from the sample. This reduced the sample to $271$ galaxies. The \verb|ellipse| isophote fitting method can fail in the case of highly inclined galaxies with extended, spheroidal stellar components, which cause $e$ to be measured as less than $0.5$ for outer isophotes. Three such galaxies were rejected after being identified through the spurious detection of an extremely strong bar by GALFIT, the visual identification of a diffuse stellar component, and the equal position angle of disk and stellar components. A preliminary application of our decomposition procedure to SDSS images revealed that reliable parameter measurements could not be obtained by GALFIT fainter than a magnitude of $\sim$$15.6$. As such, the upper $r_{\mathrm{petro}}$ limit of the sample was lowered from $16.7$ to $15.6$, decreasing the sample to $169$ galaxies.

Finally, a morphological selection was applied to the sample. This involved initially using the morphological classifications of \citet{Dressler80}. 
Representative samples of elliptical (E), S0 and spiral galaxies were constructed using SDSS DR8 $r$-band data. Visual classifications were then performed, referring to these representative samples and cross-checking with other morphological classifications \citep{Michard08} where possible. After our type classifications, a clear decision between E and S0 was not possible for a small number of fainter E/S0s; although strongly-barred S0s are easily identifiable, it is more difficult to distinguish between unbarred S0s and ellipticals. As such, there may be a few unidentified unbarred S0s not included in the sample, and the S0 bar fractions quoted in this study may therefore be upper limits. This morphological selection leaves $70$ lenticular galaxies. After the discarding of a further six galaxies due to contamination by adjacent bright  stars or companion galaxies, our final central sample size was reduced to $64$ galaxies.

\begin{figure}
\centering
\includegraphics[width=0.5\textwidth]{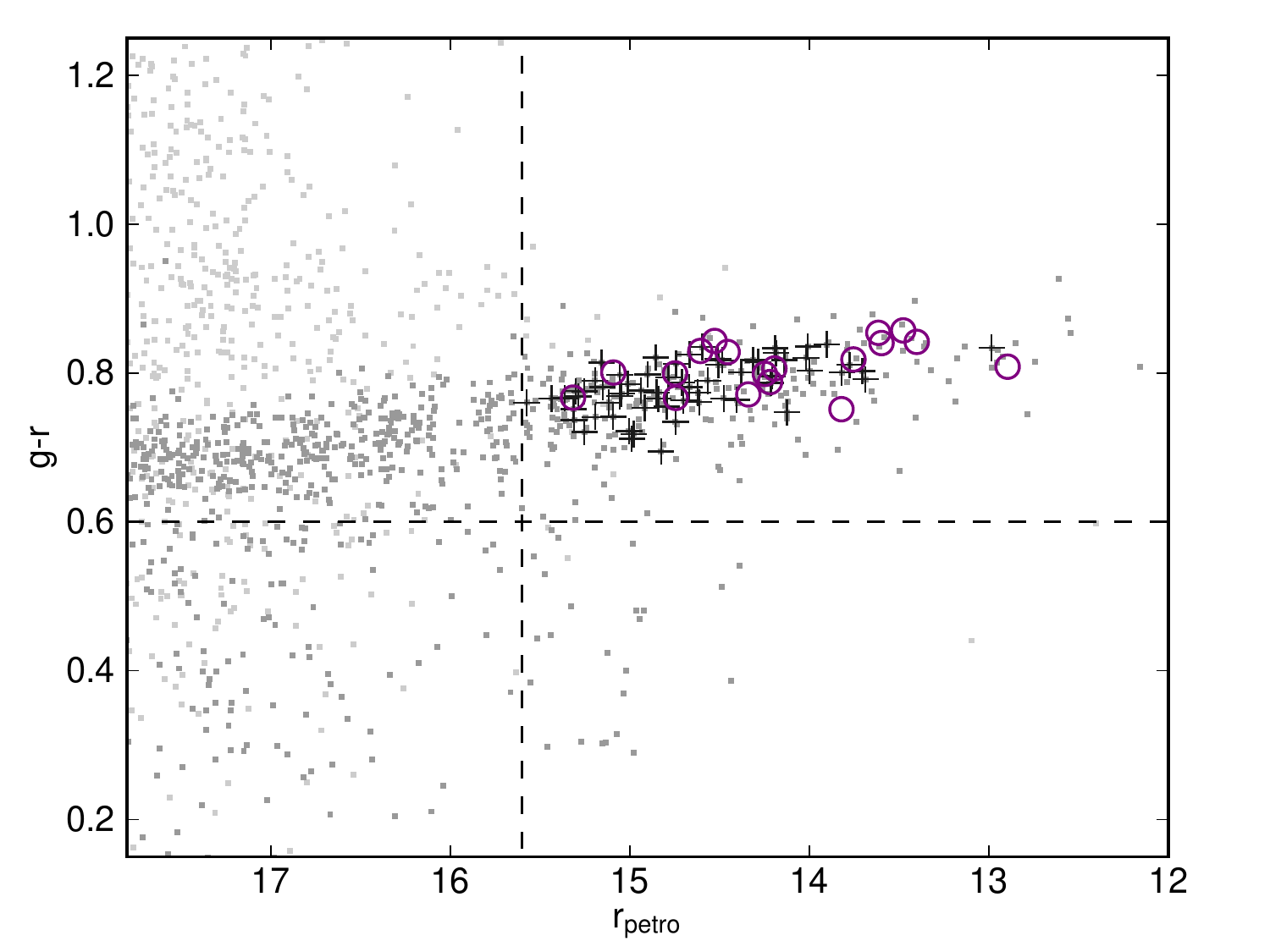}
\caption{Colour-magnitude distributions for our central 
and outskirts 
Coma cluster samples (black crosses and purple circles, respectively). For comparison, SDSS galaxies within a similar spatial region to our samples ($193.3^{\circ}<\mathrm{RA}<196.6^{\circ}$, $26.6^{\circ}<\mathrm{Dec}<29.4^{\circ}$) are shown, with dark grey and light gray squares indicating cluster members and non-members, respectively. For these comparison galaxies, cluster membership is based on the spectroscopic redshift range $0.010<z<0.037$. The dashed lines show the colour and magnitude cuts used to define our final sample; $g-r>0.6$ and $r_{\mathrm{petro}}<15.6$, respectively.}
\label{g-r_rmag}
\end{figure}

A cluster `outskirts' sample was selected with the main purpose of acting as a control sample for environment investigations. These galaxies were spectroscopically confirmed as associated with the Coma cluster in the same way as the main central sample, and selected between projected cluster radii of $2.3^{\circ}$ ($4.0$~Mpc) and $8.0^{\circ}$ ($13.9$~Mpc). 
Magnitude, inclination and Hubble type cuts as described above for the central sample, along with rejections due to contamination, resulted in an outskirts sample of $19$ S0s. 
Unlike the central sample, these outskirts galaxies lie well outside the virial radius of Coma and are unlikely to have visited the central region or experienced significant cluster interactions. 
Our samples lie on the red sequence in colour-magnitude space, as shown in Fig. \ref{g-r_rmag}.
A full list of the galaxies used in our analysis is given in Table \ref{galaxies_table}. The median point spread function (PSF) for our samples is $1.1$~arcsec, corresponding to a physical scale of $\sim$$0.5$~kpc at the distance of Coma.

\section{Results}
\label{Results}

\subsection{Example Decompositions}
\label{example_fits}

\begin{figure*}
\begin{minipage}{0.033\textwidth}
\hspace{1 mm}
\end{minipage}
\begin{minipage}{0.25\textwidth}
\includegraphics[width=\textwidth]{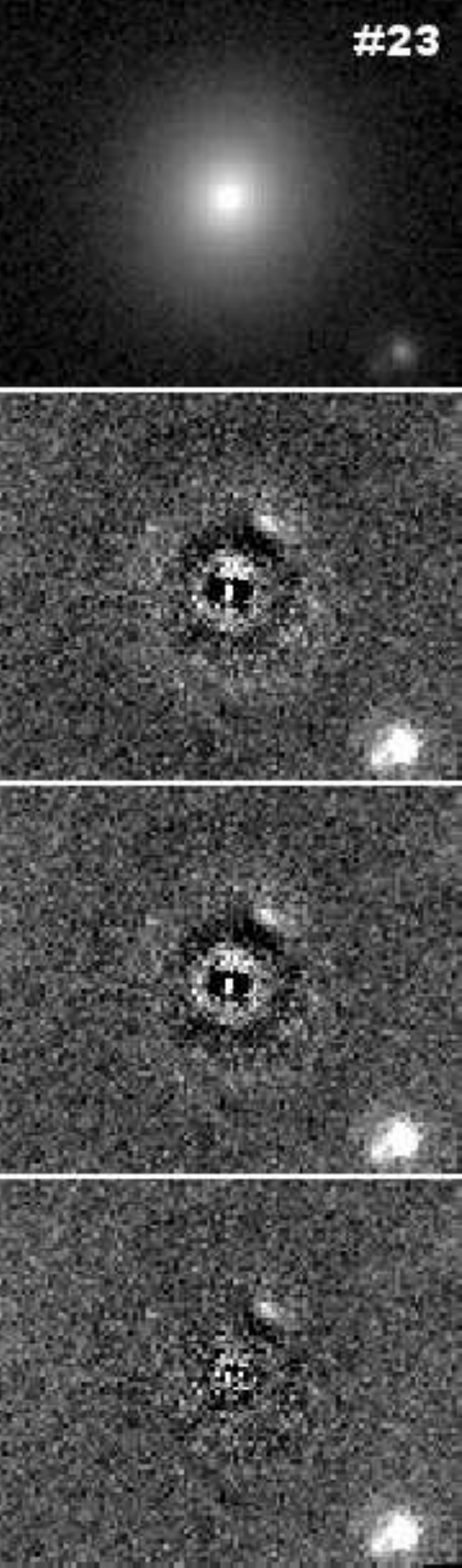}
\end{minipage}
\begin{minipage}{0.077\textwidth}
\includegraphics[width=\textwidth]{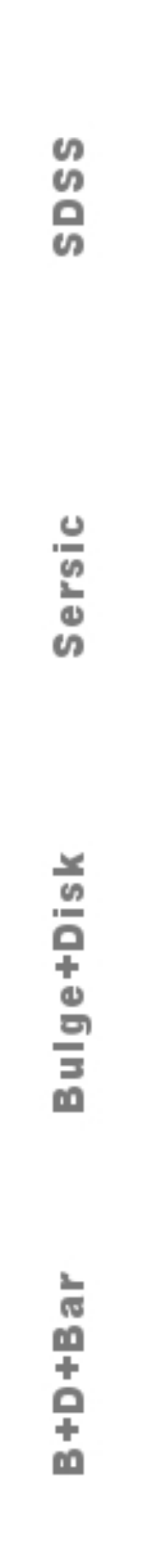}
\end{minipage}
\begin{minipage}{0.25\textwidth}
\includegraphics[width=\textwidth]{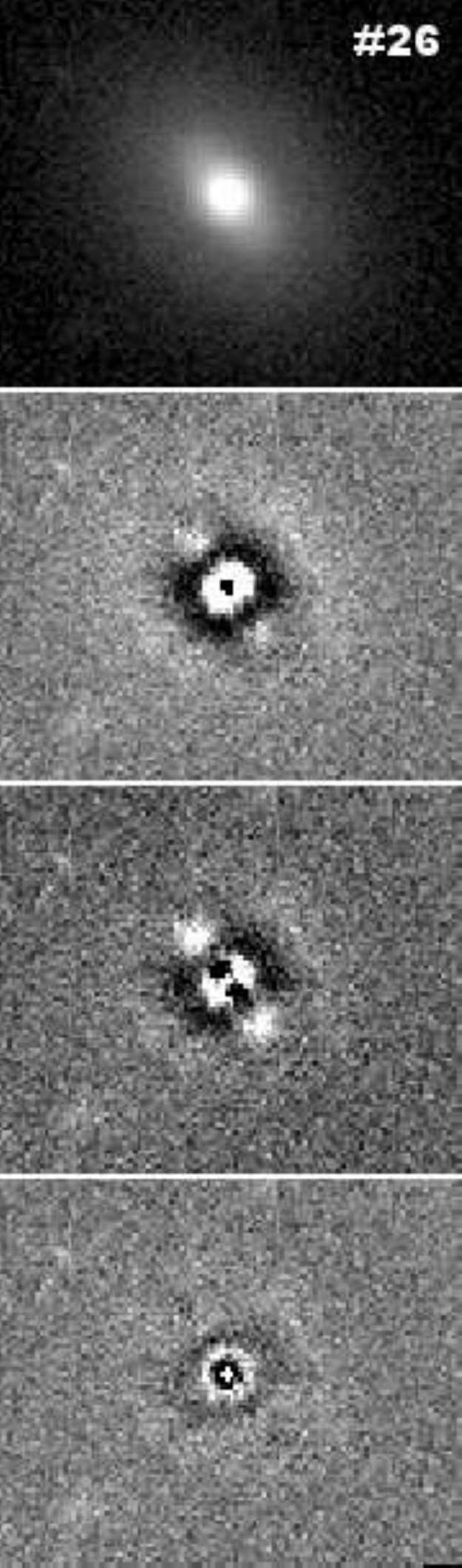}
\end{minipage}
\begin{minipage}{0.077\textwidth}
\includegraphics[width=\textwidth]{23_26_39_words}
\end{minipage}
\begin{minipage}{0.25\textwidth}
\includegraphics[width=\textwidth]{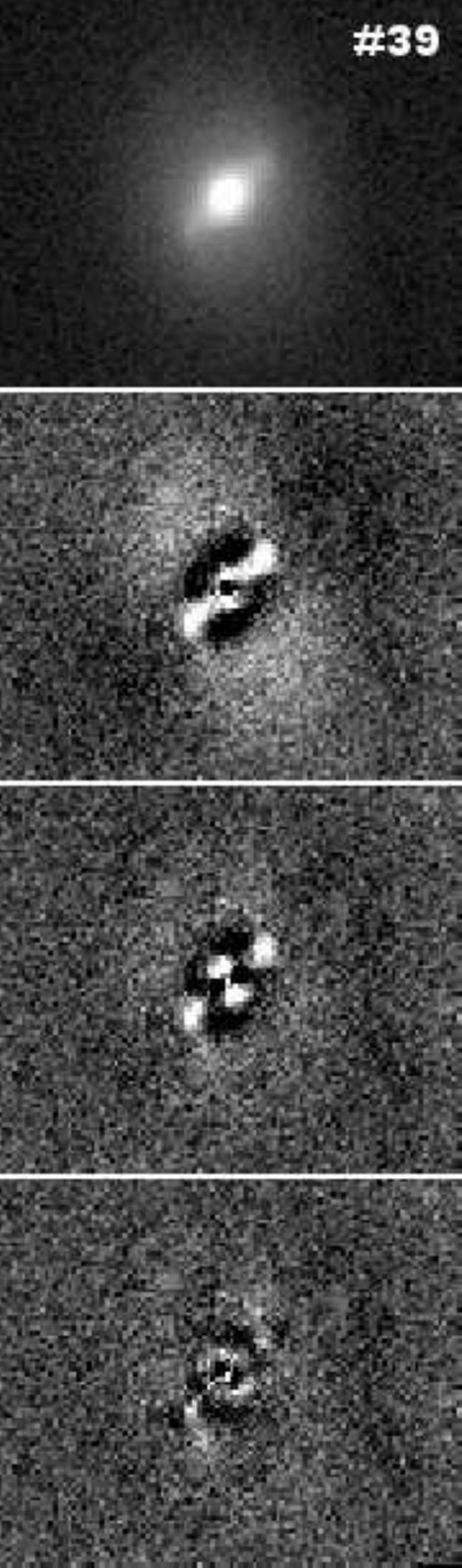}
\end{minipage}
\begin{minipage}{0.33\textwidth}
\centering
\includegraphics[width=\textwidth]{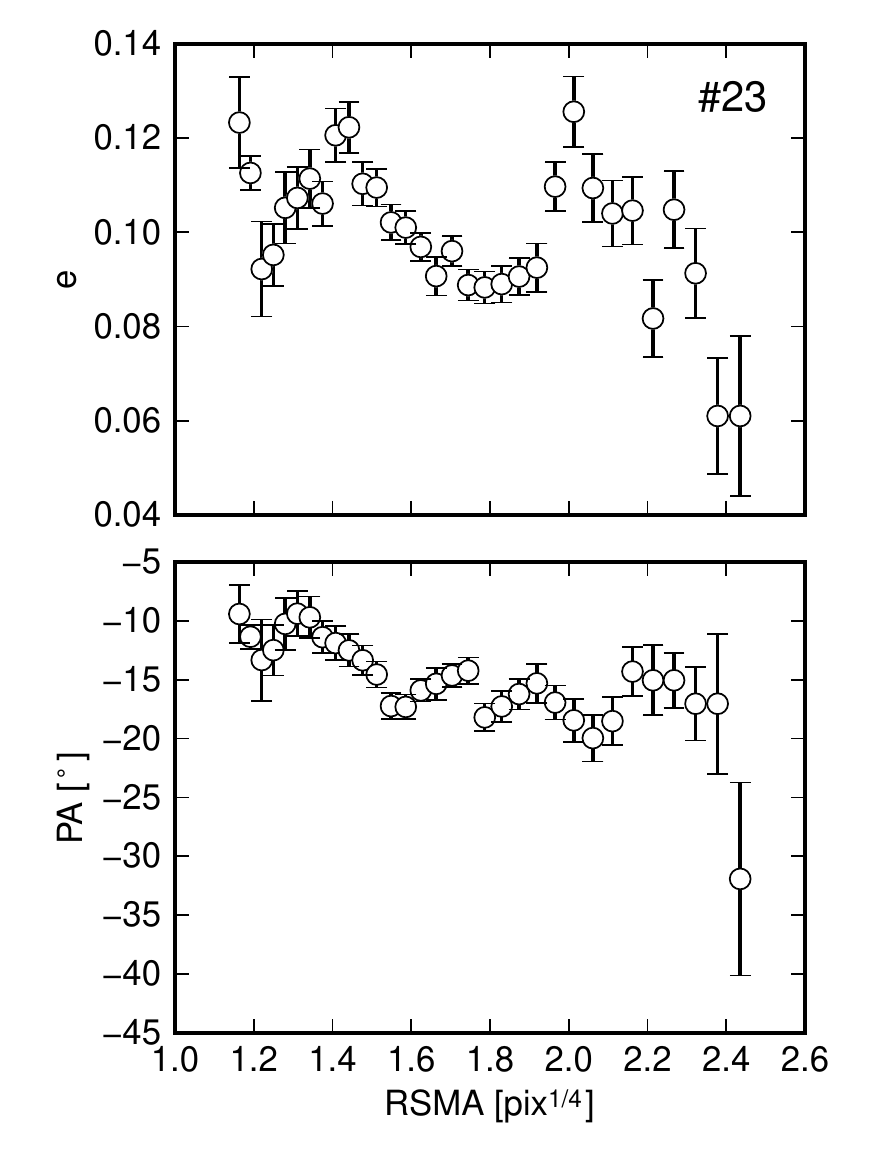}
\end{minipage}\hfill
\begin{minipage}{0.33\textwidth}
\centering
\includegraphics[width=\textwidth]{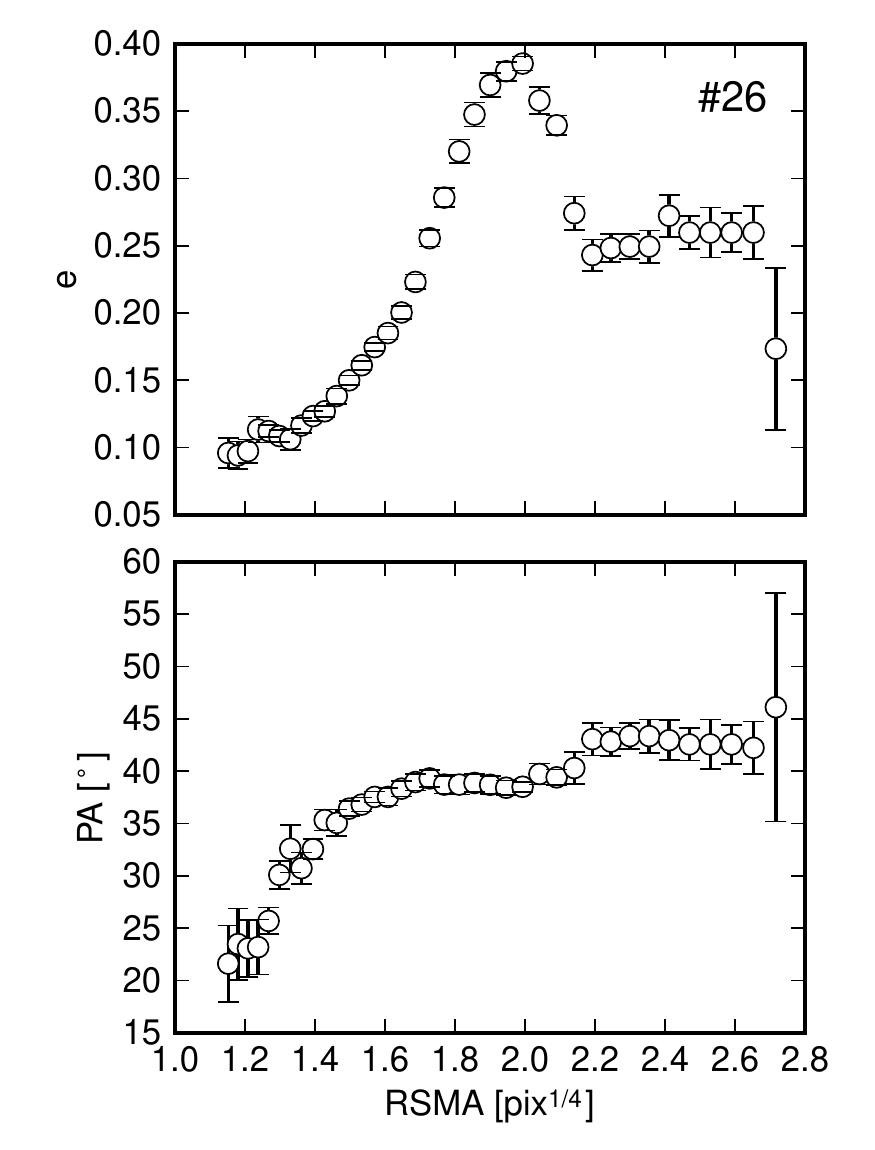}
\end{minipage}\hfill
\begin{minipage}{0.33\textwidth}
\centering
\includegraphics[width=\textwidth]{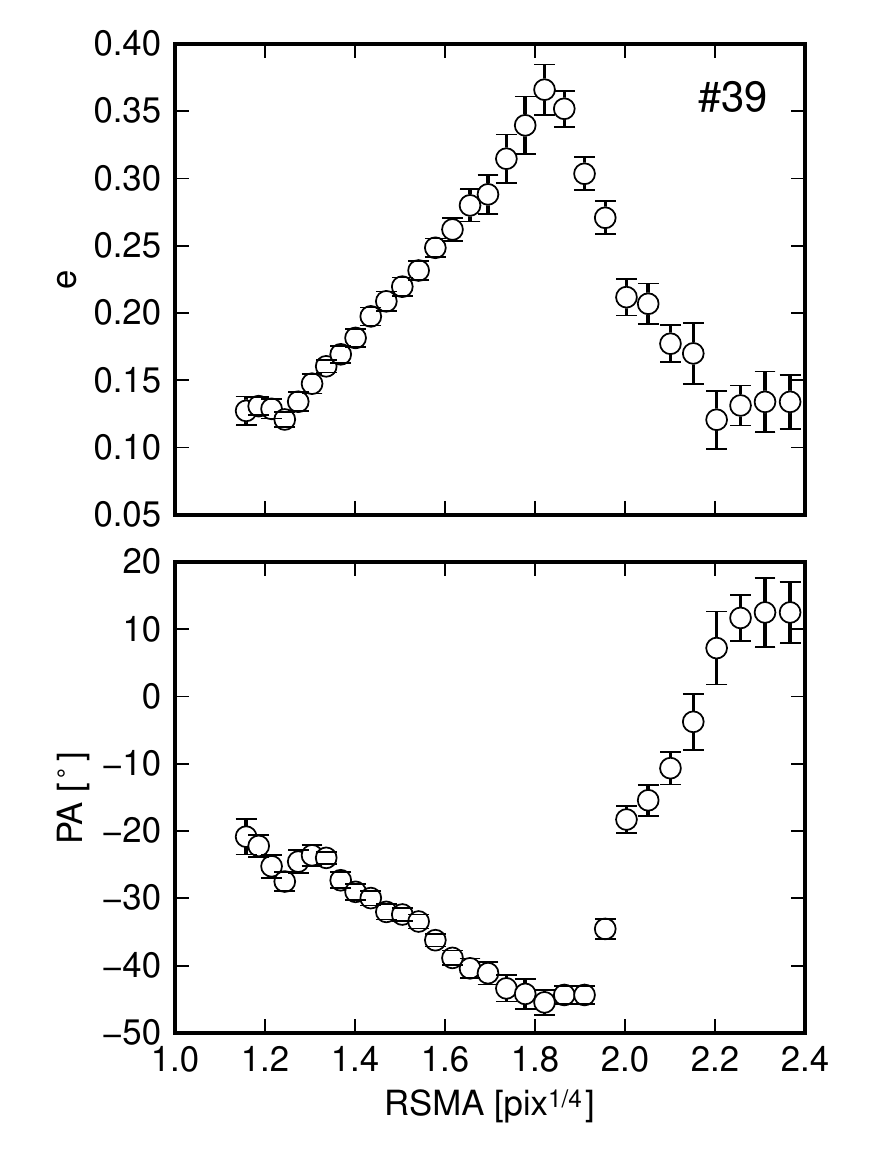}
\end{minipage}
\caption{Top panels: Structural decomposition results for an
  unbarred ($\#23$), a barred ($\#26$) and a strongly-barred ($\#39$)
  S0. Shown are SDSS $r$-band images and residual images for the three
  stages of our decomposition procedure; a single S\'ersic fit, a bulge+disk
  fit, and a bulge+disk+bar fit ($B$+$D$+$Bar$). The residual images are obtained when
  the GALFIT model is subtracted from the original SDSS image. Bottom panels:
  Isophotal analyses for the three S0s. These ellipse fitting results show the
  variation of isophote ellipticity ($e$) and position angle (PA) with
  isophote semi-major axis length (RSMA).}
\label{3_case_studies_images}
\end{figure*}

To illustrate our bulge+disk+bar decompositions we present detailed results for three example S0s; one unbarred ($\#23$), one barred ($\#26$), and one strongly-barred ($\#39$).
The three-stage image residuals and \verb|ellipse| results for these galaxies are shown in Fig. \ref{3_case_studies_images}.
Corresponding best-fit structural parameters are detailed in Table \ref{116_013_030_table}. 
We classify galaxy $\#39$ as barred (SB0) because it satisfies the bar
detection criteria in Section \ref{bar_criteria}. A strong bar
signature is observed in the image residuals that is removed when a
bar is added to the GALFIT model, evident both through visual
inspection and quantitatively with $\Delta \mathrm{RFF}=1.9\%$. This is
unsurprising as a strong bar signature is visually apparent in the
original SDSS image, and through isophotal analysis with
\verb|ellipse|. Galaxy $\#23$ does not satisfy all three of our bar detection criteria and
is subsequently classified as unbarred (S0). More interesting cases
are those of less strongly barred galaxies such as $\#26$, which satisfies our bar detection criteria and is therefore classified as SB0. This galaxy is classified as unbarred by
\citet{Dressler80}. Morphological classifications and decomposition results for the full central and outskirts samples are given in Tables \ref{galaxies_table} and \ref{params_table}, respectively.

\begin{table*}
\begin{minipage}[l]{14.0cm}
\caption{Best-fit parameters for the S\'ersic, bulge+disk and
  bulge+disk+bar fitting stages of the S0 galaxies in Fig. \ref{3_case_studies_images}. \label{116_013_030_table}}
\end{minipage}
\begin{tabular}{l c c | r c r r r r r} 

\hline\hline

\multicolumn{1}{c}{Gal.} &  
\multicolumn{1}{c}{SDSS DR8 ID} &  
\multicolumn{1}{c}{Type} &  
\multicolumn{1}{c}{$\mathrm{RFF}$} &  
\multicolumn{1}{c}{Component} &  
\multicolumn{1}{c}{$Flux/T$} &  
\multicolumn{1}{c}{$r_{e}$ [pix]} &  
\multicolumn{1}{c}{$n$} &  
\multicolumn{1}{c}{$b/a$} &  
\multicolumn{1}{c}{PA [$^{\circ}$]} \\

\multicolumn{1}{c}{(1)} &  
\multicolumn{1}{c}{(2)} & 
\multicolumn{1}{c}{(3)}  &
\multicolumn{1}{c}{(4)}  &
\multicolumn{1}{c}{(5)} & 
\multicolumn{1}{c}{(6)} &
\multicolumn{1}{c}{(7)} & 
\multicolumn{1}{c}{(8)} &  
\multicolumn{1}{c}{(9)} & 
\multicolumn{1}{c}{(10)} \\

\hline\hline
$\#23$ & $1237667443511590950$ & S0 & $0.8\%$ & S\'ersic & $1.000$ & $11.2$ & $3.02$ & $0.895$ & $-15.0$ \\
\cline{4-10}
 & &  & $0.8\%$ & bulge & $0.745$ & $7.7$ & $2.60$ & $0.880$ & $-14.9$ \\
 & &  &  & disk & $0.255$ & $18.6$ & $1.00$ & $0.968$ & $-16.6$ \\
\cline{4-10} 
 & &  & $0.2\%$ & bulge & $0.222$ & $2.5$ & $1.79$ & $0.819$ & $-13.5$ \\
 & &  &  & disk & $0.662$ & $15.0$ & $1.00$ & $0.906$ & $-15.3$ \\
 & &  &  & bar & $0.116$ & $5.2$ & $0.39$ & $0.953$ & $-19.9$ \\
\hline\hline
$\#26$ & $1237667443511787541$ & SB0 & $5.2\%$ & S\'ersic & $1.000$ & $20.5$ & $7.87$ & $0.683$ & $40.0$ \\
\cline{4-10}
 & &  & $2.6\%$ & bulge & $0.489$ & $3.5$ & $2.58$ & $0.694$ & $39.4$ \\
 & &  &  & disk & $0.512$ & $30.4$ & $1.00$ & $0.740$ & $42.5$ \\
\cline{4-10}
 & &  & $1.0\%$ & bulge & $0.363$ & $2.2$ & $1.68$ & $0.822$ & $41.1$ \\
 & &  &  & disk & $0.559$ & $26.5$ & $1.00$ & $0.726$ & $42.2$ \\
 & &  &  & bar & $0.077$ & $9.8$ & $0.34$ & $0.385$ & $37.8$ \\
\hline\hline
$\#39$	& $1237667444048724176$ & SB0	& $5.3\%$	& S\'ersic & $1.000$ & $35.7$	& $6.68$	& $0.731$	& $-33.9$	\\
\cline{4-10}
	& & 	& $3.1\%$	& bulge	& $0.419$ & $4.7$	& $2.67$	& $0.603$	& $-39.8$	\\
	& & 	& 	& disk	& $0.589$ & $22.3$	& $1.00$	& $0.864$	& $12.5$		\\
\cline{4-10}
	& & 	& $1.2\%$	& bulge	& $0.205$ & $1.9$	& $1.48$	& $0.783$	& $-16.7$	\\
	& & 	& 	& disk	& $0.675$ & $20.1$	& $1.00$	& $0.805$	& $1.8$		\\
	& & 	& 	& bar	& $0.121$ & $8.2$	& $0.48$	& $0.338$	& $-47.7$	\\
\hline
\end{tabular}
\begin{minipage}[l]{14.0cm}
\small
(1) Galaxy ID for this study.
(2) SDSS DR8 object ID.
(3) Hubble type as determined in this study.
(4) Residual flux fractions for the three fitting stages, as defined in Equation \ref{RFF_equation}.
(5) Model components for each fitting stage.
(6) Light fraction.
(7) Effective radius.
(8) S\'ersic index.
(9) Axial ratio.
(10) Position angle.
\end{minipage}
\end{table*}

\subsection{Comparison with Previous Studies}
\citet{Dressler80} morphologically classified galaxies in the Coma cluster, recording which S0s were barred. For the seven S0s in our sample typed by Dressler as barred, we also detect bars.
However, for the $30$ S0s typed by Dressler as unbarred, we 
find that $20$ have evidence for a bar. 
The larger dynamic range of CCD data allows structures to be detected that
may not have been apparent on the 103a-O photographic plates used by Dressler.

\citet{Marinova12} use ellipse fitting of images from the \emph{HST} ACS Coma Cluster survey \citep{Carter08} to detect bars in S0s. Of the $13$ S0s Marinova et al. classify as barred, eight make it into our sample and we agree in every case that a bar is present.

\subsection{Bar Fractions}
\label{bar_fractions}

The bar fraction ($f_{\mathrm{bar}}$) is defined as the fraction of
disk galaxies (exclusively S0 galaxies in this study) which host bars. Bar fractions for our central and outskirts cluster samples are given in Table \ref{fbar_table}.
\begin{table}
\caption{S0 bar fractions for Coma. \label{fbar_table}}
\centering
\begin{tabular}{c c c} \hline\hline
Study & Detection Method & S0 Bar Fraction ($f_{\mathrm{bar}}$) \\
\hline
\multicolumn{3}{c}{core sample ($R_{\rm proj}<0.37$~Mpc)} \\
\hline
This study	&	B+D+Bar	&	$85^{+6}_{-10}\%$ ($17/20$)	\\
This study	&	ellipse (relaxed)	&	$60^{+10}_{-11}\%$ ($12/20$)	\\
\citet{Marinova12}	&	ellipse (relaxed)	&	$65^{+10}_{-11}\%$$^{a}$ ($13/20$)	\\
\hline
\multicolumn{3}{c}{central sample ($R_{\rm proj}<2.5$~Mpc)} \\
\hline
This study	&	B+D+Bar	&	$72^{+5}_{-6}\%$ ($46/64$)	\\
This study	&	ellipse (relaxed)	&	$48^{+6}_{-6}\%$ ($31/64$)	\\
This study	&	ellipse (strict)	&	$41^{+6}_{-6}\%$ ($26/64$)	\\
\hline
\multicolumn{3}{c}{outskirts sample ($4<R_{\rm proj}<14$~Mpc)} \\
\hline
This study	&	B+D+Bar	&	$58^{+11}_{-11}\%$ ($11/19$)	\\
This study	&	ellipse (relaxed)	&	$32^{+11}_{-10}\%$ ($6/19$)	\\
\hline

\end{tabular}
\begin{minipage}[l]{0.47\textwidth}
\small
$^{a}$: To aid the comparison of studies we have propagated these errors using the same error analysis as for our sample, so they are not as originally published. 
\end{minipage}
\end{table}
We obtain $f_{\mathrm{bar}}=72^{+5}_{-6} \%$ for the central sample. 
Results are also included for a `core' cluster sub-sample, for S0s with $R_{\mathrm{proj}}< 0.37$~Mpc. This $R_{\mathrm{proj}}$ limit corresponds to that used by \citet{Marinova12}.
The bar fraction errors given are $1\sigma$ (i.e. $68.3\%$ confidence level)
binomial uncertainties. As discussed in Section \ref{sample
  selection}, some further uncertainty 
arises due the exclusion of a small
number of morphologically ambiguous E/S0 galaxies during sample
selection, which may have boosted
our bar fraction measurements with respect to the true values. 
There may also be a bias in the opposite direction caused by the less-than-unity bar detection efficiency of our
method, inferred from model galaxy fitting (see Section \ref{model_detection}).
For instance, correcting the central sample bar fraction for
the inferred missing SB0s yields $f_{\mathrm{bar}}=83^{+4}_{-5} \%$. 
For simplicity we have not propagated these additional uncertainties.

In order to compare our work with recent studies we have also measured
$f_{\mathrm{bar}}$ by detecting bars using the ellipse fitting of
galaxy isophotes with the IRAF task \verb|ellipse|. This has been done
with both `strict' detection criteria, where a global maximum in
ellipticity ($e$) outwards from the galaxy centre is required for a
galaxy to be considered barred, and `relaxed' criteria, where a local
maximum in $e$ suffices (the adopted criteria, which follow
  \citealt{Barazza09a} and \citealt{Marinova10}, are detailed in Section \ref{method_intro}). There are no S0s for which a bar is detected using ellipse fitting but not using bulge+disk+bar decomposition. In contrast to this, $20$ S0s have bars detected using bulge+disk+bar decomposition that are not detected using ellipse fitting with relaxed detection criteria. As a result, bar fractions obtained using ellipse fitting are considerably lower, by a factor of $\sim$$1.6$, than those obtained using bulge+disk+bar decomposition. 
In Appendix \ref{appendix} we detail why for five of the bars not detected using ellipse fitting there is a degree of uncertainty in our bar detections. Our bar fraction results are discussed in the context of other recent studies in Section \ref{Discussion}.


\subsection{Bars in S0s as a Function of Galaxy Luminosity and Colour}
\label{results_galaxy_properties}

\begin{figure}
\centering
\includegraphics[width=0.47\textwidth]{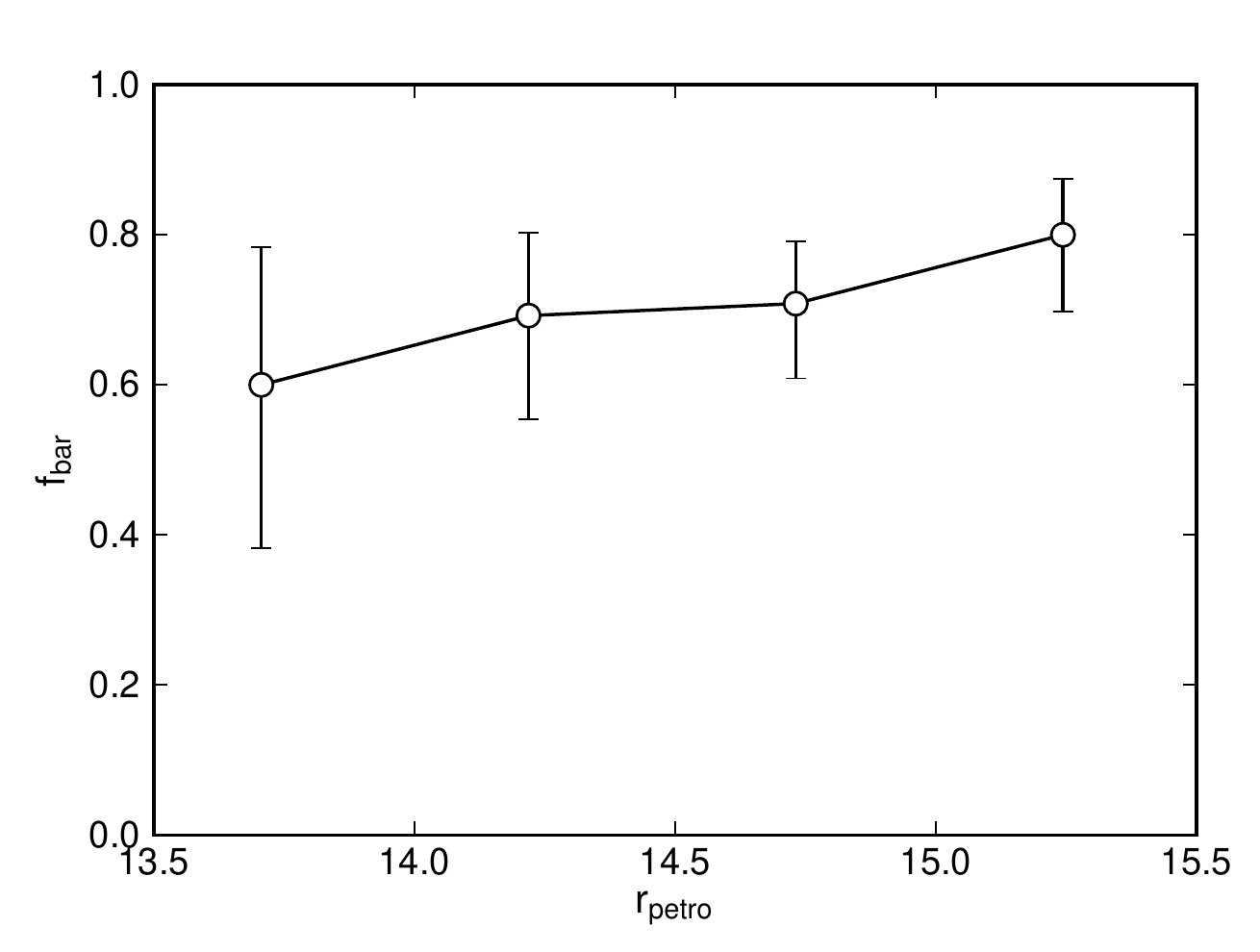}
\caption{S0 bar fraction ($f_{\mathrm{bar}}$) against $r$-band Petrosian magnitude ($r_{\mathrm{petro}}$). A weak trend is observed, but this is not statistically significant.}
\label{rmag_graph1}
\end{figure}

\begin{figure*}
\centering
\includegraphics[width=1\textwidth]{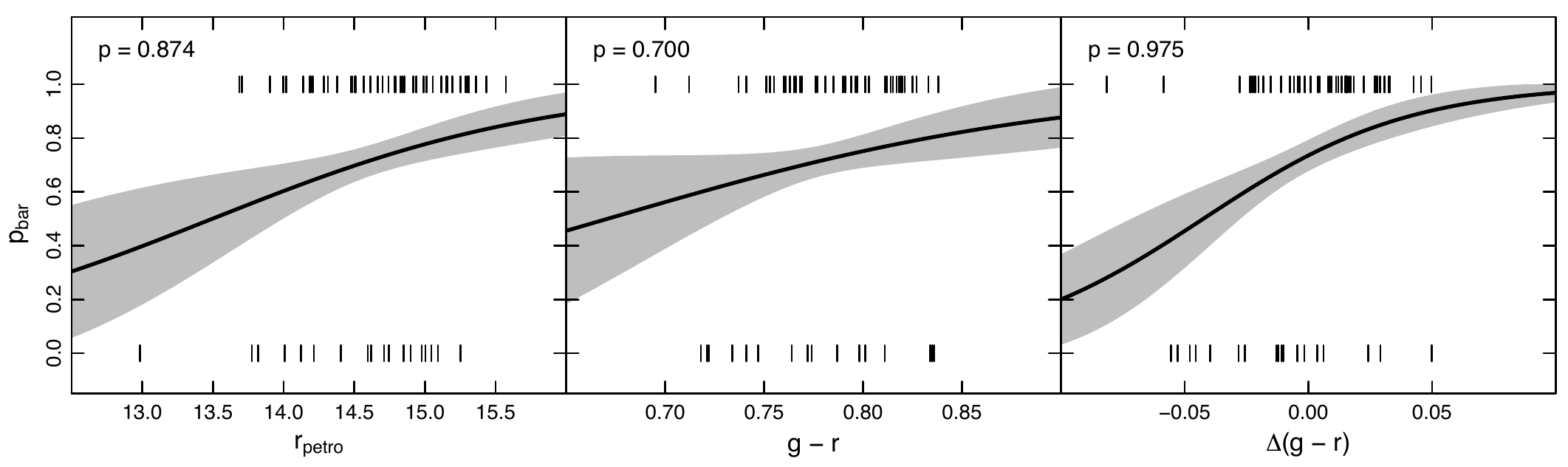}
\caption{Logistic regression results for the correlation of bar probability ($p_{\rm bar}$) with magnitude ($r_{\rm petro}$), colour ($g-r$) and offset from the colour-magnitude relation ($\Delta(g-r)$). Barred and unbarred galaxies are shown as ticks at $p_{\rm bar}$\,=\,1 and $p_{\rm bar}$\,=\,0, respectively. The shaded regions indicate 1$\sigma$ errors in the predicted mean $p_{\rm bar}$. We note the $p$-value for rejecting the hypothesis of no correlation in each panel. The correlation of $p_{\rm bar}$ with $\Delta(g-r)$ is significant at the $>2\sigma$ level.}
\label{LRA_CMR}
\end{figure*}

Here we present our results regarding bar dependence on
galaxy luminosity and colour.
In Fig. \ref{rmag_graph1}, the bar fraction
($f_{\mathrm{bar}}$) of our sample is shown as a function of $r$-band Petrosian
magnitude ($r_{\rm petro}$). Although we find no significant correlation,
we cannot rule out variation within a range of $\sim$$\pm
20\%$. This large uncertainty
is due to the small number of galaxies  
in each bin. In an alternative approach we have applied a logistic regression analysis \citep[e.g.,][]{Hosmer00} to quantify the
correlation between the probability of a galaxy hosting a bar, $p_{\rm
  bar}$, and $r_{\rm petro}$. This has also been carried out for $g-r$
colour. The logistic regression analysis explicitly accounts for the
dichotomous nature of the dependent variable (barred vs. unbarred), and
does not require the binning of data. There is a marginally-significant ($p=0.874$)
increase in $p_{\rm bar}$ towards fainter magnitudes, as shown in the left panel of Fig.
\ref{LRA_CMR}. The correlation
with colour is not significant (Fig. \ref{LRA_CMR}, centre panel). 

Combining the colour and magnitude information, we find that $\Delta(g-r)$, defined as the offset in colour of a galaxy from the mean colour-magnitude relation, is the best predictor of whether an S0 will host a bar. The correlation of $p_{\rm bar}$ with $\Delta(g-r)$ (Fig. \ref{LRA_CMR}, right panel) is significant at the $>2\sigma$ level, i.e. galaxies which are redder than average for their luminosity are more likely to host bars.
Since the $g-r$ values associated with this result were obtained from SDSS `model'
magnitudes, they reflect the global colour of the galaxies. To
probe colours within the inner regions of galaxies, the logistic
regression analysis was repeated using SDSS $7.43$ and
$3.00$~arcsec aperture colours. For these apertures, the correlation of $p_{\rm bar}$ with $\Delta(g-r)$ is
significant at the $\sim$$3\sigma$ and $\sim$$2.5\sigma$ levels, respectively.

\begin{figure}
\centering
\begin{minipage}{0.47\textwidth}
\centering
\includegraphics[width=\textwidth]{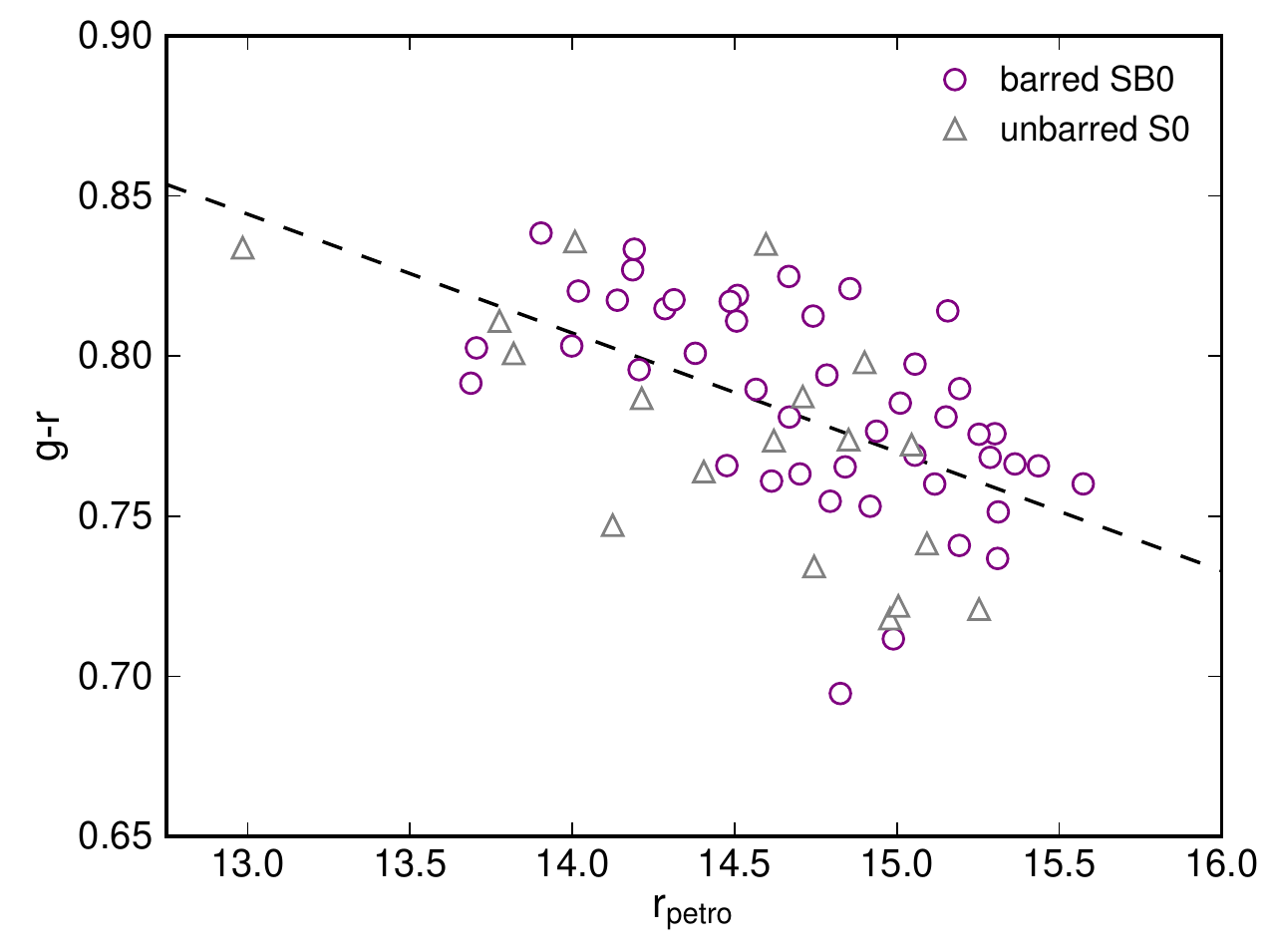}
\end{minipage}
\caption{Colour-magnitude diagram for the central cluster sample. SB0s and S0s are represented by purple circles and grey triangles, respectively. The distribution of SB0s appears, on average, redder than that of S0s. The dashed line shows the mean colour-magnitude relation.}
\label{gr_vs_rmag}
\end{figure}

\begin{figure}
\centering
\begin{minipage}{0.46\textwidth}
\centering
\includegraphics[width=\textwidth]{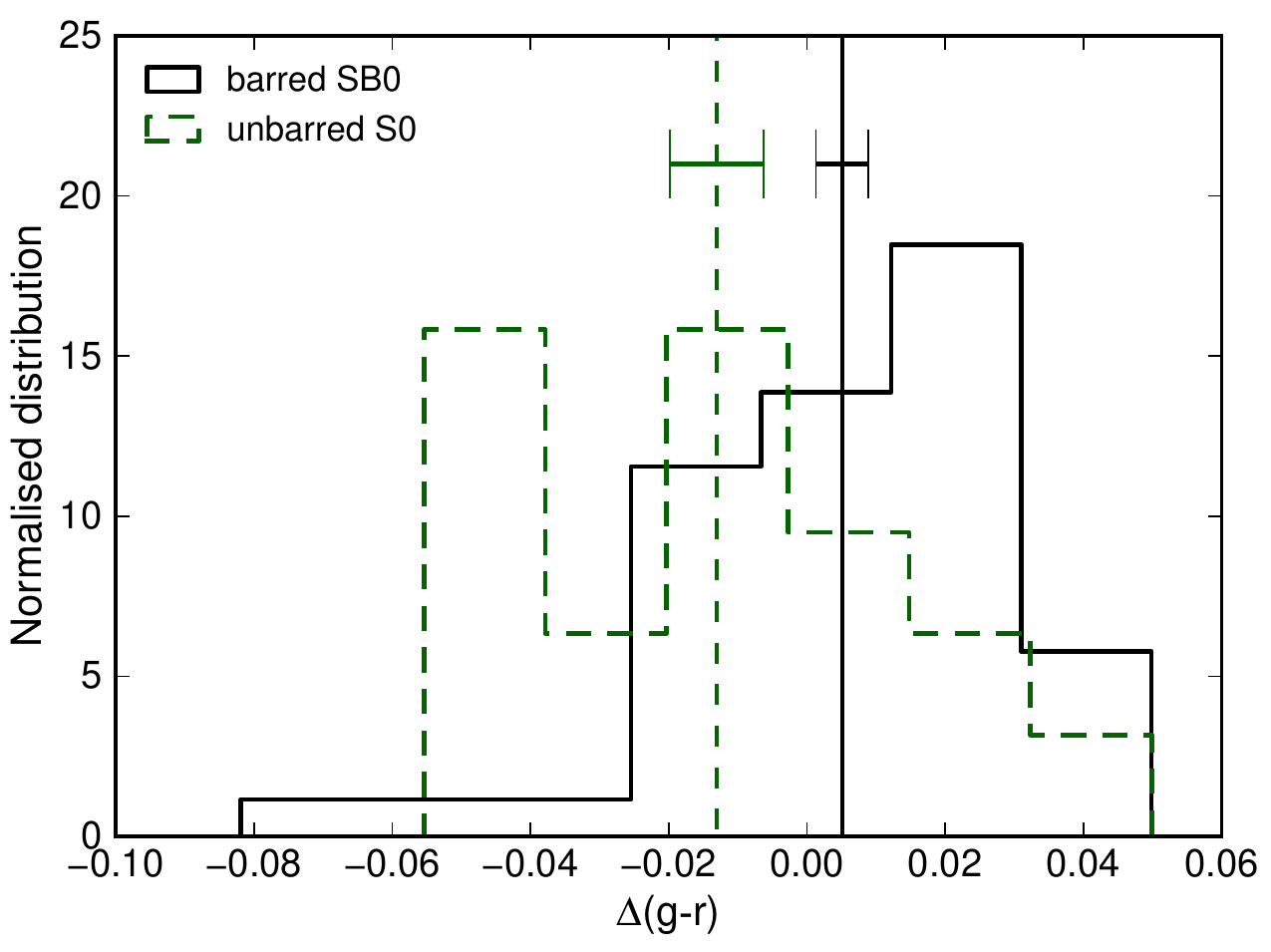}
\end{minipage}
\caption{Normalised distributions of $g-r$ offsets from the colour-magnitude relation ($\Delta(g-r)$) for barred (solid black) and unbarred (dashed green) S0s in the central sample. Vertical lines indicate mean values, the standard errors of which are overlaid. The systematic offset of the mean values is significant at the $>2\sigma$ level. If real, this offset indicates that for a given magnitude bars are more likely to be found in redder galaxies.}
\label{c_hists}
\end{figure}

The result that galaxies redder than average for their luminosity are 
more likely to host bars is readily apparent when the colour-magnitude diagram is
considered (Fig. \ref{gr_vs_rmag}).
Histograms of $\Delta(g-r)$ for barred and unbarred S0s are shown in
Fig. \ref{c_hists}. The systematic offset between the mean values of
the two distributions, $0.018\pm0.008$, is significant at the
$2.3\sigma$ level; a similar significance to
that obtained from the regression analysis above. Repeating this test
using $7.43$ and $3.00$~arcsec aperture colours yields
offset significances of $3.0\sigma$ and $2.8\sigma$, respectively.
The consistency of the above results implies that the correlation of
bars and $\Delta(g-r)$ is a global effect, and not attributed
to a specific region of the galaxy, e.g. the bar or bulge.

\begin{figure}
\centering
\includegraphics[width=0.47\textwidth]{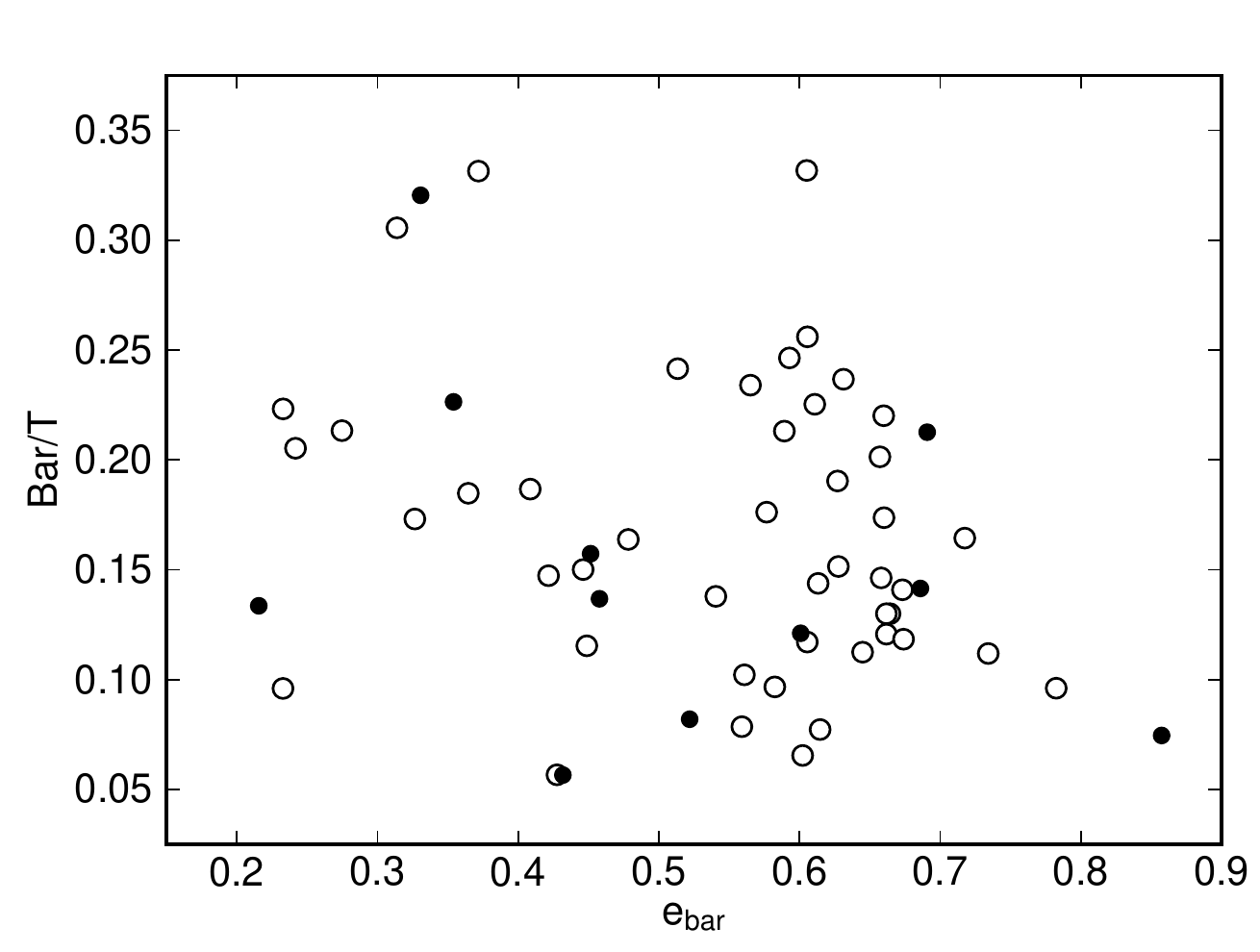}
\caption{Bar ellipticity ($e_{\mathrm{bar}}$) against bar light fraction ($Bar/T$) for our central and outskirts cluster samples (empty and solid circles, respectively), as measured using bulge+disk+bar decomposition. The general scatter indicates that the two parameters are not closely correlated.}
\label{LF_vs_e}
\end{figure}

We now consider the strength of detected bars. Here, and in Sections \ref{results_age_metallicity} and \ref{results_environment}, we use $\Phi_{\mathrm{bar}}=e_{\mathrm{bar}}\times Bar/T$ as a quantitative bar strength parameter, following \citet{Weinzirl09b}. As shown in Fig. \ref{LF_vs_e}, our measured $e_{\mathrm{bar}}$ and $Bar/T$ parameters are not correlated, 
which confirms that the two are independent measures of bar strength.

\begin{figure}
\centering
\includegraphics[width=0.47\textwidth]{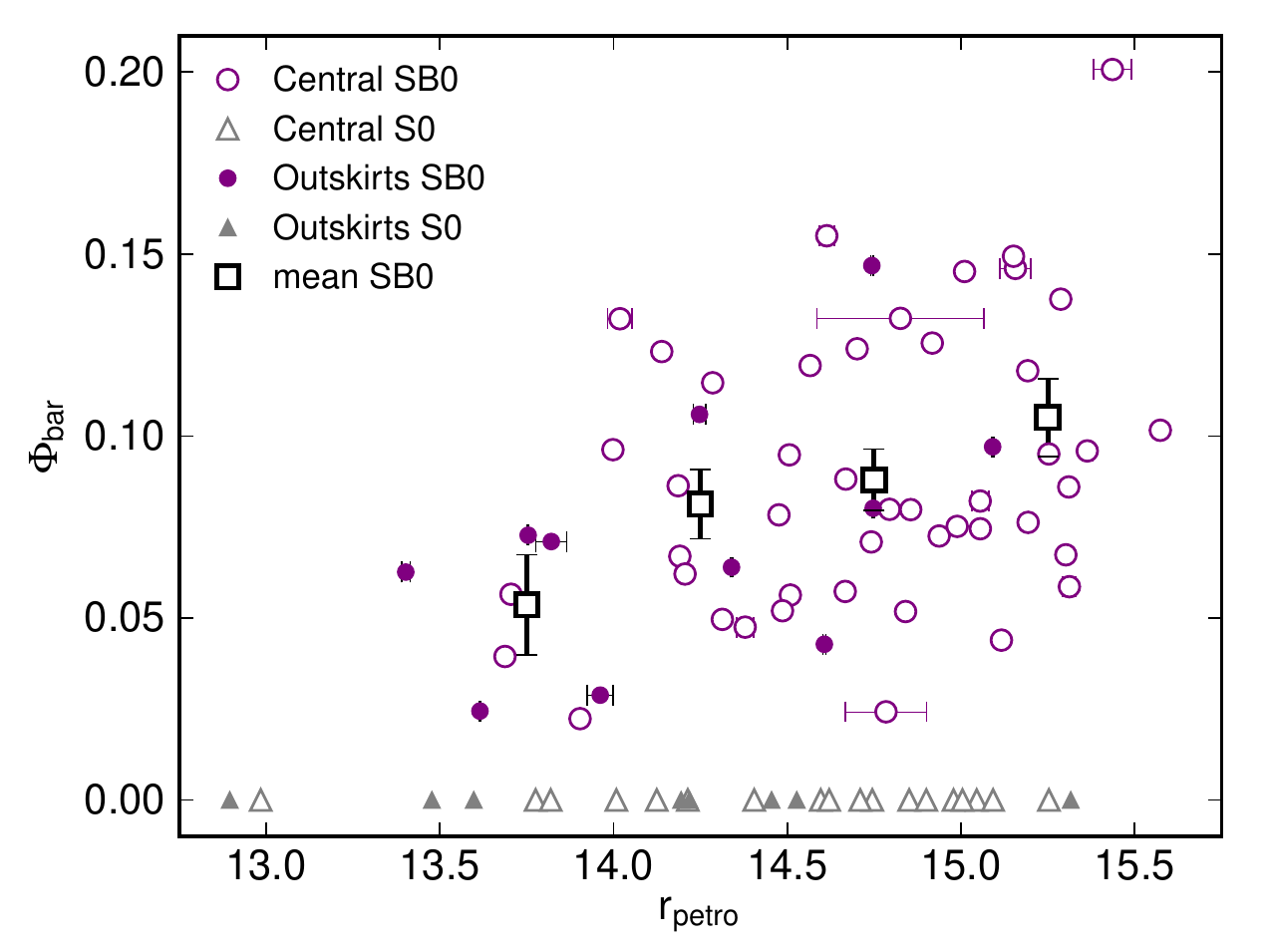}
\caption{Bar strength ($\Phi_{\mathrm{bar}}$) against $r$-band Petrosian magnitude ($r_{\mathrm{petro}}$) for the central and outskirts samples (empty and solid shapes, respectively). SB0s and S0s are represented by purple circles and grey triangles, respectively. Mean values of $\Phi_{\mathrm{bar}}$ for the SB0s are over-plotted as black squares, the error bars of which are calculated using the standard deviation of $\Phi_{\mathrm{bar}}$ within each $r_{\mathrm{petro}}$ bin. An increase in $\Phi_{\mathrm{bar}}$ towards fainter magnitudes is observed.}
\label{rmag_graph2}
\end{figure}

While we measure an increase in $\Phi_{\mathrm{bar}}$ towards fainter magnitudes (Fig. \ref{rmag_graph2}),  
the likely systematic biases need to be considered. For instance, the GALFIT decomposition procedure may overestimate
$\Phi_{\mathrm{bar}}$ for fainter galaxies, or may not be able to
detect weaker bars in fainter galaxies. To address these
issues, we apply the decomposition procedure described in Section
\ref{fitting_routine} to $25$ artificial SB0 galaxies with
$\Phi_{\mathrm{bar}}=0.03$ and $r_{\mathrm{petro}}\sim15.3$, values
which correspond to the region of concern in Fig.
\ref{rmag_graph2}. Bars were successfully measured in all
$25$. Following our artificial galaxy analysis in Section
\ref{magnitude_limit}, Fit/Model values were calculated for the GALFIT
parameters of each galaxy. A mean value of
$\mathrm{Fit/Model}=1.05\pm0.05$ was obtained for
$\Phi_{\mathrm{bar}}$. Since this is consistent with unity, we
conclude that our decomposition procedure does not significantly
overestimate $\Phi_{\mathrm{bar}}$ for fainter galaxies.

\subsection{Bars in S0s as a Function of Stellar Age and Metallicity}
\label{results_age_metallicity}

To investigate variations in bar properties with stellar age and metallicity, we use the stellar populations measurements of \citet{Smith12}. These were derived from SDSS spectra which sample an aperture diameter of $3$~arcsec.

The barred and unbarred S0s in our sample, for which there are stellar populations data available, occupy similar regions of age-metallicity space. A KS-test was performed to determine whether the stellar age distributions of the barred and unbarred S0s differ significantly. This yielded (KS: $p=0.34, D=0.3$), where $p$ is the p-value of the hypothesis test and $D$ is the maximum difference between the  cumulative distribution functions. KS-tests were also performed for the Fe/H and Mg/Fe distributions, yielding (KS: $p=0.80, D=0.2$) and (KS: $p=0.35, D=0.3$), respectively. In all three cases a null hypothesis cannot be rejected; the results are consistent with equivalent central stellar populations for barred and unbarred S0s.

\begin{figure*}
\centering
\includegraphics[width=1\textwidth]{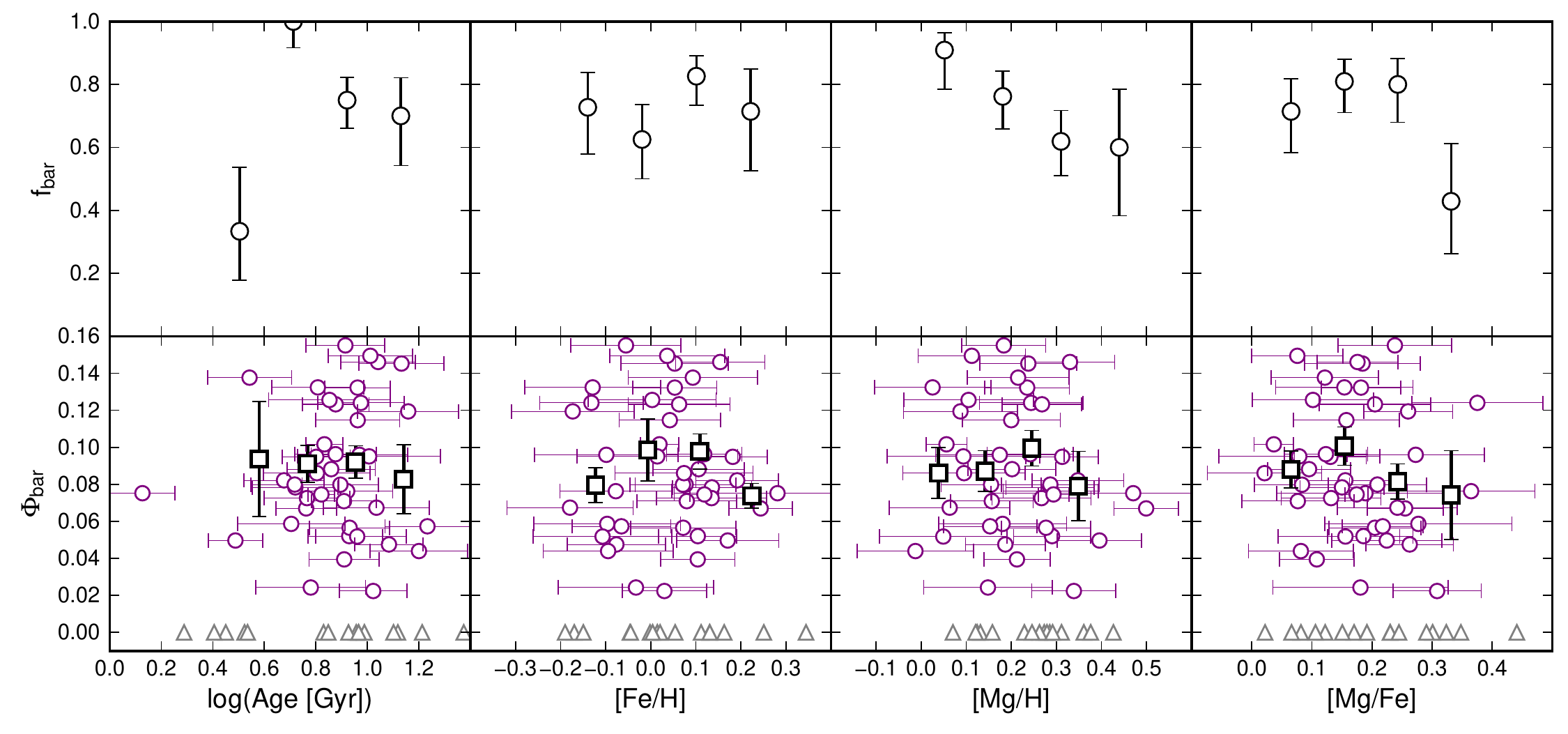}
\caption{$f_{\mathrm{bar}}$ (top row) and $\Phi_{\mathrm{bar}}$ (bottom row) against SSP-equivalent stellar age in units of Gyr ($\log(\rm{Age})$), iron (Fe/H) and magnesium (Mg/H) abundances, and abundance ratio (Mg/Fe). Labelling for the bottom row follows that of Fig. \ref{rmag_graph2}. No evidence is found for strong correlations between $f_{\mathrm{bar}}$ or $\Phi_{\mathrm{bar}}$ and stellar age or metallicity.}
\label{agemet_full}
\end{figure*}

\begin{figure*}
\centering
\includegraphics[width=1\textwidth]{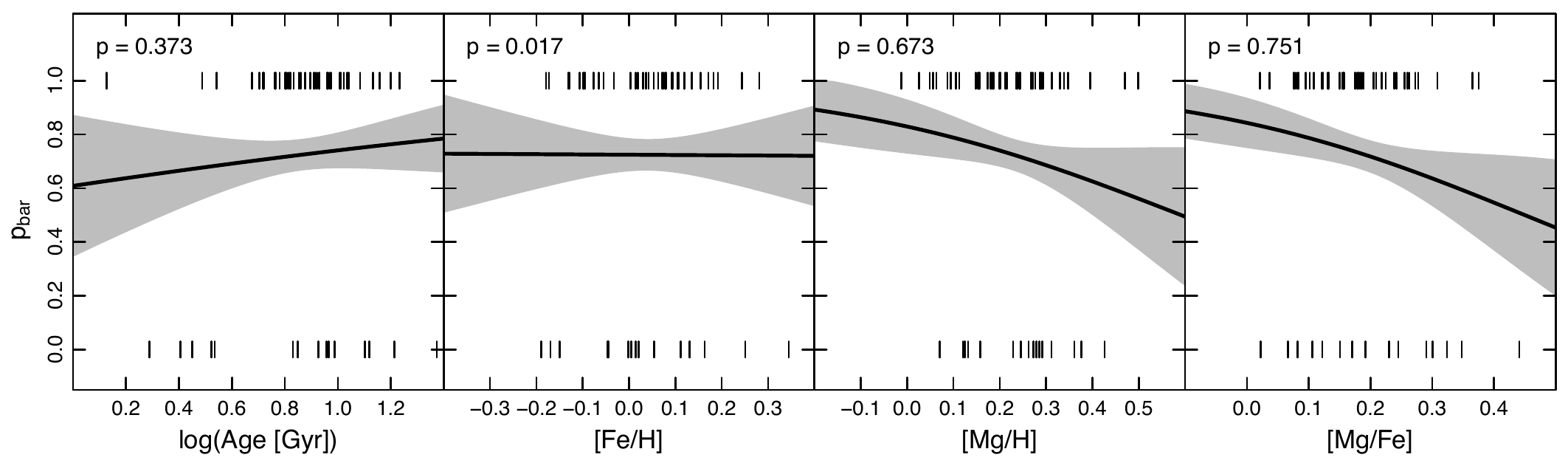}
\caption{Bar probability ($p_{\mathrm{bar}}$) against SSP-equivalent stellar age in units of Gyr ($\log(\rm{Age})$), iron (Fe/H) and magnesium (Mg/H) abundances, and abundance ratio (Mg/Fe). The panels are analogous to those of Fig. \ref{LRA_CMR}. There is no significant correlation between the presence of a bar and stellar age or metallicity.}
\label{LRA_agemet}
\end{figure*}

Trends with various stellar population parameters are investigated in Fig. \ref{agemet_full}. 
There is no clear evidence for correlations between $f_{\mathrm{bar}}$
or $\Phi_{\mathrm{bar}}$ and stellar age or metallicity. 
Again we have applied a logistic regression analysis \citep[e.g.,][]{Hosmer00} to quantify the correlations.
Results for the dependence of $p_{\rm bar}$ on age, Fe/H, Mg/H and Mg/Fe are shown in Fig. \ref{LRA_agemet}. For all four of these parameters, the analysis confirms the impression given by Fig. \ref{agemet_full}, i.e. that no significant correlations are present.

\subsection{Bars in S0s as a Function of Environment}
\label{results_environment}

\begin{figure}
\centering
\includegraphics[width=0.47\textwidth]{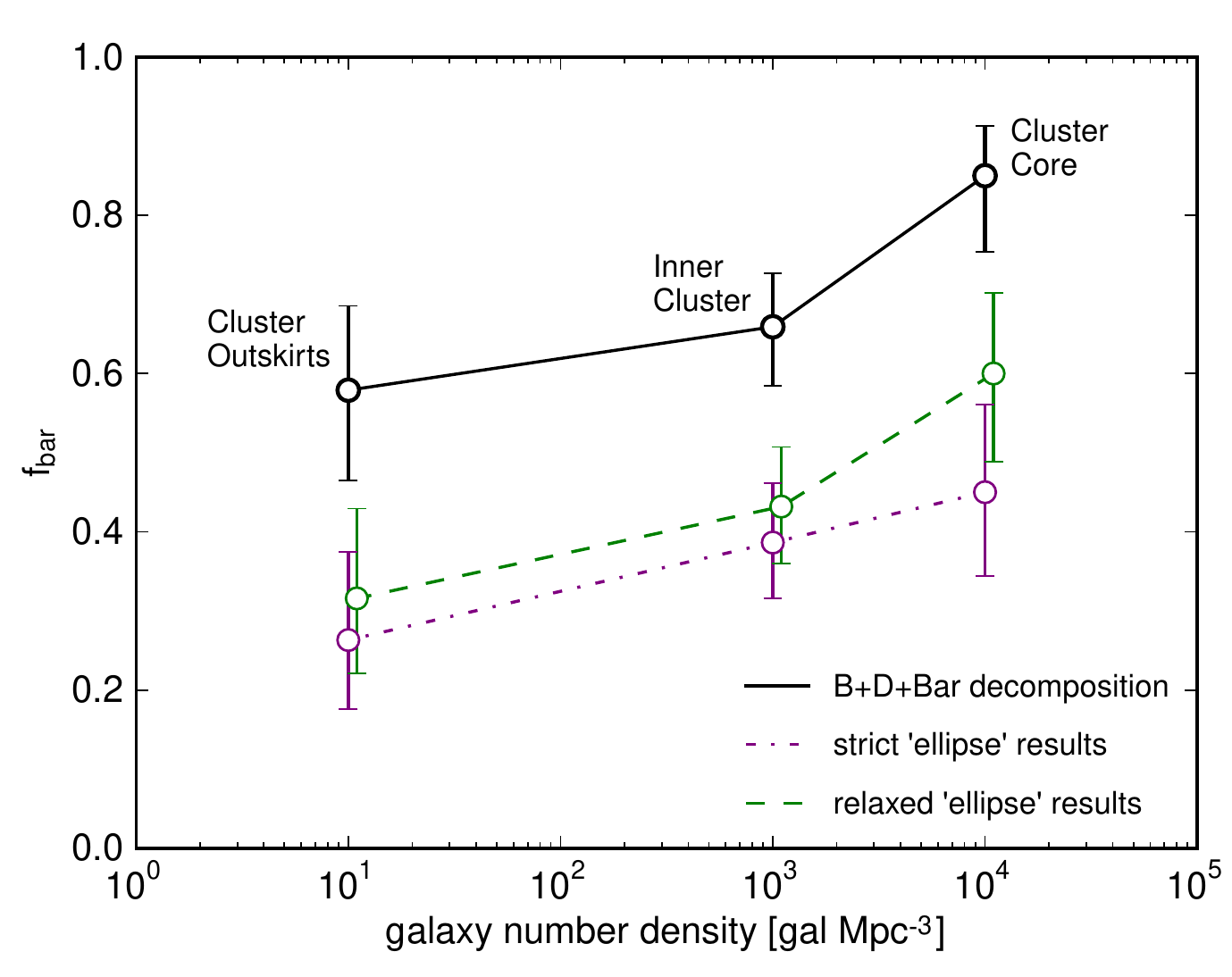}
\caption{Bar fraction ($f_{\mathrm{bar}}$) against galaxy number density for S0s in Coma. The results shown were obtained using three bar detection methods; bulge+disk+bar decomposition (solid line), ellipse fitting with `strict' bar detection criteria (dashed-dotted line), and ellipse fitting with `relaxed' criteria (dashed line). `Outskirts', `Inner' and `Core' refer to our outskirts sample, $0.37$--$2.5$~Mpc sub-sample and core sub-sample, respectively.}
\label{fbar_enviro}
\end{figure}

\begin{figure}
\centering
\includegraphics[width=0.47\textwidth]{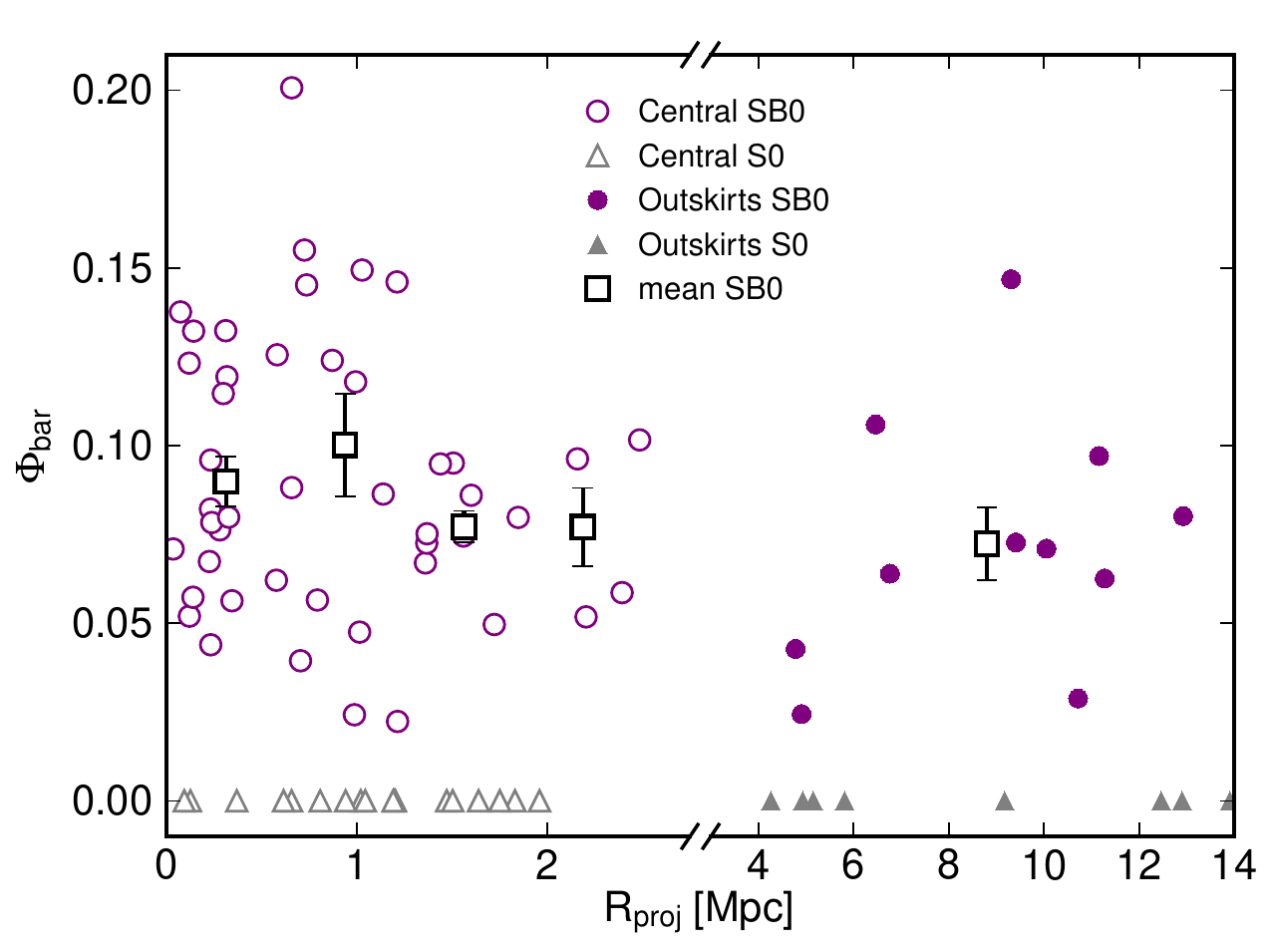}
\caption{Bar strength ($\Phi_{\mathrm{bar}}$) against projected cluster radius ($R_{\mathrm{proj}}$). Labelling follows that of Fig. \ref{rmag_graph2}. A weak trend is observed, but this is not statistically significant.}
\label{phi_vs_Rproj}
\end{figure}

To explore the variation of bars between environments of significantly
different densities, our central sample is divided into a `core'
sub-sample, for galaxies with $R_{\mathrm{proj}}<0.37$~Mpc, and
a `$0.37$--$2.5$~Mpc' sub-sample. The galaxy number densities
($n$) of the Coma core, the $0.37$--$2.5$~Mpc and the outskirts environments are
$n\sim10000$ gal Mpc$^{-3}$, $n\sim1000$ gal Mpc$^{-3}$ and $n\sim10$
gal Mpc$^{-3}$, respectively \citep{The86,Marinova12}. We measure
similar bar fractions for the outskirts sample and the
$0.37$--$2.5$~Mpc sub-sample of
$f_{\mathrm{bar}}=58^{+11}_{-11} \%$ (N~$=11/19$) and
$f_{\mathrm{bar}}=66^{+7}_{-7} \%$ (N~$=29/44$), respectively, and a considerably
larger fraction for the core sub-sample of
$f_{\mathrm{bar}}=85^{+6}_{-10} \%$ (N~$=17/20$). These $f_{\mathrm{bar}}$ results,
along with equivalent results we obtained using ellipse fitting, are
plotted against galaxy number density in Fig. \ref{fbar_enviro}. The
observed increase in $f_{\mathrm{bar}}$ for the cluster core is at the $\sim$$1.5\sigma$ significance level.
Bar strength ($\Phi_{\mathrm{bar}}$) is plotted as a function of $R_{\mathrm{proj}}$ in Fig. \ref{phi_vs_Rproj}. 
While a weak trend with environment is observed, this is also of low significance.

\section{Discussion}
\label{Discussion}

We have developed techniques for bar detection and carried out a detailed analysis of bars in S0s in the Coma cluster, including their correlations with various galaxy properties. Here we discuss the results of our analysis in the context of other studies. 

Locally, the optical bar fraction ($f_{\mathrm{bar}}$) is around $\sim$$50\%$ when measured across all disk galaxy types (S0 to Im) and environments \citep{Reese07,Barazza08,Aguerri09a}. This rises to around two-thirds when NIR images are included \citep{Knapen00,Eskridge00,Marinova07,Menendez-Delmestre07}. 
The S0 bar fractions we measure are significantly higher than both these general values, and those reported in previous studies of Coma \citep[e.g.,][]{Thompson81,
Marinova10,Marinova12}. 
For instance, in the most recent comparable study of bars in S0s in Coma, performed using images from the \emph{HST} ACS Coma Cluster survey \citep{Carter08}, \citet{Marinova12} use the ellipse fitting method to measure a bar fraction of $f_{\mathrm{bar}}=65^{+10}_{-11}\%$ in the cluster core. 
Our S0 bar fraction for the same region is considerably higher, $f_{\mathrm{bar}}=85^{+6}_{-10}\%$. We obtain a similar fraction to Marinova et al. when using their ellipse method to detect bars, $f_{\mathrm{bar}}=60^{+10}_{-11}\%$. 
Considering all S0s studied in our work, $20$ have bars detected using bulge+disk+bar decomposition that are not detected using the ellipse method.
This suggests either that our method is more efficient at detecting bars or that it erroneously detects bars where there are not any. As detailed in Appendix \ref{appendix}, for $15$ of these $20$ S0s we are confident that bars have been detected, but for the remaining five there is still, for a variety of reasons, a degree of uncertainty.

Our results provide strong evidence that S0s which are redder than average for their luminosity are more likely to host bars. 
Previous studies of disk galaxies, which included S0s,
have found a higher bar fraction in galaxies with
redder global optical colours \citep[e.g.][]{Masters11}.
While the observed bar-colour dependence may be driven
by stellar populations, our results find
no significant trend with either central stellar age or metallicity.
However, there are significantly larger uncertainties in the stellar populations measurements with respect to the general scatter of the data than for SDSS colours. As such, any bar-stellar populations correlation may be difficult to measure.
Nevertheless, for a reliable comparison of the two results we have considered the galaxy regions being probed.
While the stellar populations data is for the bulge-dominated central $1.4$~kpc region, our measured correlation between bar probability and colour offset appears to be a global effect. Importantly, the correlation is still significant at the $\sim$$2.5$-$2.8\sigma$ level when only colours from the central $1.4$~kpc are considered. 
We conclude that although a significant bar-colour dependence is observed, interpretations about the driving factor, be it stellar ages, metallicities or some combination of both, are limited by the uncertainties in the spectroscopic data available.

We find weak evidence that fainter S0s are more likely to host bars 
in agreement with the results of \citet{Laurikainen13}, who use a large sample of early-types, and \citet{Barway11}, who also study S0s in clusters. 
Furthermore, we measure an increase in bar strength towards fainter luminosities. These effects may be understood by considering different evolutionary histories for bright and faint S0s in clusters. The idea that S0s are transformed spirals that have lost their gas supply is favoured due to their position on the red sequence, lack of molecular gas, 
and the observation of an abundance of blue spirals in clusters above $z\sim0.2$ but not in the local universe \citep[e.g.,][]{Butcher84}. When a spiral galaxy moves at high speed through the intra-cluster medium (ICM), cold interstellar gas in its disk can be lost to the environment. 
This process is only significant in the densest environments such as Coma, where there is observational evidence of stripping in the form of gas tails \citep{Smith10}. If a barred spiral is subject to such stripping, stellar features such as the bar may remain intact. 
Therefore, a possible explanation of the luminosity correlations is that
fainter S0s were preferentially formed through removal of gas from
spirals at late epochs, whereas brighter S0s formed through another
process (e.g., mergers) which tended to erase pre-existing bars.
This argument relies on the assumption that spirals host stronger bars and significantly more bars than lenticulars, as observed by \citet{Buta08} and \citet{Laurikainen09}, respectively.

We find that the central stellar age distributions of barred and unbarred Coma S0s do not differ significantly, and hence find no evidence from stellar ages that bars are linked with bulge formation. For comparison, \citet{Coelho11} use a large statistical sample and find significantly different distributions for barred and unbarred galaxies. 
We note that their study includes disk galaxies up to very late-types, while we study specifically S0s, and any bar-driven bulge formation is likely to depend on type-specific properties such as gas availability. \citet{Perez11}, who focus mostly on early-types, obtain results consistent with our work, i.e. no strong evidence for differences in the stellar age distributions of barred and unbarred galaxies. 

Measurements of high central stellar metallicities in barred galaxies may be explained by bar-enhanced SFRs, due to bar-driven gas inflow during bulge formation. \citet{Perez11} obtain such a result and conclude that bars may be long-lived structures, closely linked with bulge formation. Our results disagree with this scenario; we find that barred and unbarred S0s are consistent with having similar metallicity distributions. This implies that bars in Coma have not had a significant impact on the chemical evolution of their host galaxies, at least in the galactic centres. Possible explanations may be that the bars are too young \citep[$\sim$$10^{7}\mathrm{yr}$,][]{Considere00} to have had an effect, or simply that bar-driven gas inflows do not significantly affect the chemical evolution of galaxies.

Numerical simulations indicate that high speed tidal encounters in the dense cores of clusters may be effective at inducing bars in disk galaxies, despite the short timescales over which they act \citep{Romano08,Aguerri09b}.
We measure an increase in $f_{\mathrm{bar}}$ between the low density outer regions of Coma, and its high density core (Fig. \ref{fbar_enviro}). Although this agrees with similar measurements in other studies \citep{Thompson81,Barazza09a,Marinova12}, like these results, ours is of low significance. 
It thus remains difficult to rule out the possibility that high speed encounters do not induce bars, or that the combination of low gas contents and tidal heating, which hinder bar instabilities, rules out the tidal induction of bars in clusters. Our results support the picture that external processes do not strongly impact bar evolution.

\section{Conclusions}
\label{Conclusions}

We have used SDSS DR8 $r$-band images to study bars in S0s in the Coma cluster. We have analysed a sample of $64$ central cluster, and $19$ outskirts members. With artificial galaxy images we have demonstrated that bulge+disk+bar decomposition is an effective bar detection method, determined a magnitude limit for the successful measurement of bars, 
and introduced a quantitative bar detection parameter $\Delta \mathrm{RFF}$. 
Our main conclusions are:
\begin{itemize}

\item The overall optical bar fraction of our central cluster sample is $72^{+5}_{-6} \%$. This high value is due to the bulge+disk+bar decomposition method being more sensitive to the presence of bars than other techniques.

\item We find strong evidence that for a given luminosity, barred S0s are redder in $(g-r)$ colour than unbarred S0s by $0.02$~mag. 

\item We measure an increase in the frequency and strength of bars towards fainter luminosities, which may be linked to different evolutionary histories for bright and faint S0s in Coma. 

\item Neither the stellar age nor metallicity distributions of our barred and unbarred S0s differ significantly. We find no clear evidence for bars playing an important role in bulge building or the chemical enrichment of central regions.

\item We measure a higher bar fraction in the dense core of Coma compared to lower density outer regions, but this is at a low significance level. 
Bars in Coma have at most a weak dependence on cluster-centric radius.

\end{itemize}

\section*{Acknowledgements}
We are grateful to Jacob Head for useful discussions. G.B.L.\ received financial support from the Science and Technology Facilities Council (STFC). R.J.S.\ was supported by STFC Rolling
Grant PP/C501568/1 `Extragalactic Astronomy and Cosmology at Durham 2008–2013'.

This paper is based on observations collected as part of the Sloan Digital Sky Survey (SDSS). Funding for the SDSS has been provided by the Alfred P. Sloan Foundation, the Participating Institutions, the National Science Foundation, the U.S. Department of Energy, the National Aeronautics and Space Administration, the Japanese Monbukagakusho, the Max Planck Society, and the Higher Education Funding Council for England. The SDSS Web Site is http://www.sdss.org/. The SDSS is managed by the Astrophysical Research Consortium for the Participating Institutions. The Participating Institutions are the American Museum of Natural History, Astrophysical Institute Potsdam, University of Basel, University of Cambridge, Case Western Reserve University, University of Chicago, Drexel University, Fermilab, the Institute for Advanced Study, the Japan Participation Group, Johns Hopkins University, the Joint Institute for Nuclear Astrophysics, the Kavli Institute for Particle Astrophysics and Cosmology, the Korean Scientist Group, the Chinese Academy of Sciences (LAMOST), Los Alamos National Laboratory, the Max-Planck-Institute for Astronomy (MPIA), the Max-Planck-Institute for Astrophysics (MPA), New Mexico State University, Ohio State University, University of Pittsburgh, University of Portsmouth, Princeton University, the United States Naval Observatory, and the University of Washington.

\appendix

\section{Supplementary Information for Individual Objects}
\label{appendix}

For $20$ of our analysed S0s we detected bars using bulge+disk+bar decomposition, but not using \verb|ellipse| isophote fitting (for both strict and relaxed criteria).
For $15$ of these, we are confident that bars have been detected via bulge+disk+bar decomposition due to a combination of: convincing GALFIT decomposition parameters, high $\Delta \mathrm{RFF}$ values, convincing bar signatures in the bulge+disk residuals in terms of shape/pattern, and the visual identification of bars in SDSS images.
Below we briefly discuss the other five S0s for which there is still a degree of uncertainty in bar detection.

$\#05$-- The results show convincing decomposition parameters, a high $\Delta \mathrm{RFF}$ value of $3.29\%$ (see Table \ref{params_table}), and a strong bar signature in the image residuals. However, looking at the original SDSS image we were unable to come to a firm conclusion as to whether or not this galaxy is highly inclined with an extended spheroidal stellar component. 

$\#07$-- Although this galaxy satisfies all bar detection criteria, the visual change in image residuals when a bar component is added is not as significant as for other S0s for which we detect bars. 

$\#44$-- The results show convincing decomposition parameters, a high $\Delta \mathrm{RFF}$ value of $1.84\%$, and a strong bar signature in the image residuals. Additionally, sensible structural parameters can only be converged upon when a bar is included in the GALFIT model. However, looking at the original SDSS image we were unable to come to a firm conclusion as to whether or not this galaxy is highly inclined with an extended spheroidal stellar component.

$\#47$-- Although this galaxy satisfies all bar detection criteria, a combination of being at the faint end of our sample and at relatively high inclination has resulted in the image residuals bar signature being poorly defined.

$\#65$-- Here the bar detection uncertainty arises from a ring-like `bar' signature in bulge+disk residuals. However, if GALFIT is fitting a ring we would expect the bar component axial ratio to be similar to that of the disk, whereas we measure axial ratios of $0.54$ and $0.71$ for the bar and disk components, respectively. 

\counterwithin{figure}{section}
\counterwithin{table}{section}

\clearpage
\begin{table*}
\begin{minipage}[l]{15.57cm}
\caption{SDSS data and morphological classifications for the $83$ lenticular (S0) galaxies in this investigation. $\#1$-$64$ are the central sample, and $\#65$-$83$ are the outskirts sample. \label{galaxies_table}}
\end{minipage}
\begin{tabular}{lcrcccccccc}
\hline \hline
\multicolumn{1}{l}{ID} &  
\multicolumn{1}{c}{SDSS ID} & 
\multicolumn{1}{c}{$R_{\mathrm{proj}}$}  &
\multicolumn{1}{c}{$\log\sigma$}  &
\multicolumn{1}{c}{$r_{\mathrm{petro}}$} & 
\multicolumn{1}{c}{$g$-$r$} &
\multicolumn{1}{c}{Type} & 
\multicolumn{1}{c}{Type} &  
\multicolumn{1}{c}{Type} & 
\multicolumn{1}{c}{`ellipse'?} & 
\multicolumn{1}{c}{`ellipse'?} \\

\multicolumn{1}{l}{$\#$} &  
\multicolumn{1}{c}{(DR8)} & 
\multicolumn{1}{c}{Mpc}  &
\multicolumn{1}{c}{${\rm{km}\,\rm{s}^{-1}}$}  &
\multicolumn{1}{c}{mag} & 
\multicolumn{1}{c}{mag} &
\multicolumn{1}{c}{(This Study)} & 
\multicolumn{1}{c}{(D80)} &  
\multicolumn{1}{c}{(M08)} & 
\multicolumn{1}{c}{(strict)} & 
\multicolumn{1}{c}{(relaxed)} \\

\multicolumn{1}{l}{(1)} &  
\multicolumn{1}{c}{(2)} & 
\multicolumn{1}{c}{(3)}  &
\multicolumn{1}{c}{(4)}  &
\multicolumn{1}{c}{(5)} & 
\multicolumn{1}{c}{(6)} &
\multicolumn{1}{c}{(7)} & 
\multicolumn{1}{c}{(8)} &  
\multicolumn{1}{c}{(9)} & 
\multicolumn{1}{c}{(10)} & 
\multicolumn{1}{c}{(11)} \\

\hline
01 & 1237665440442089583 & 1.020 & 2.111 & 15.045 & 0.772 & S0 & S0 & - & N & N \\
02 & 1237665440442089601 & 0.941 & 1.680 & 15.253 & 0.721 & S0 & S0 & - & N & N \\
03 & 1237665440442351629 & 1.138 & 2.211 & 14.186 & 0.827 & SB0 & SB0 & - & Y & Y \\
04 & 1237667323797635139 & 0.657 & 2.272 & 13.775 & 0.811 & S0 & S0 & SB0 & N & N \\
05 & 1237667323797635239 & 0.736 & 2.164 & 15.010 & 0.785 & SB0 & S0 & Sp & N & N \\
06 & 1237667323797962933 & 0.993 & 1.803 & 15.192 & 0.741 & SB0 & SB0 & - & Y & Y \\
07 & 1237667324334374925 & 0.987 & 2.019 & 14.784 & 0.794 & SB0 & S0 & - & N & N \\
08 & 1237667324334374981 & 0.871 & 2.157 & 14.701 & 0.763 & SB0 & Ep & - & Y & Y \\
09 & 1237667324334440636 & 0.657 & 2.108 & 14.668 & 0.781 & SB0 & S0 & SBa & Y & Y \\
10 & 1237667324334440637 & 0.657 & 1.962 & 15.436 & 0.766 & SB0 & S0 & Sa & Y & Y \\
11 & 1237667324334571728 & 0.119 & 2.205 & 14.139 & 0.817 & SB0 & S0 & S.. & Y & Y \\
12 & 1237667324334571849 & 0.280 & 2.094 & 15.193 & 0.790 & SB0 & E & SB0 & N & N \\
13 & 1237667324334637140 & 0.230 & 2.073 & 15.055 & 0.769 & SB0 & S0 & S0 & N & N \\
14 & 1237667324334637189 & 0.237 & 2.085 & 14.476 & 0.766 & SB0 & SB0 & SB0 & Y & Y \\
15 & 1237667324334637285 & 0.369 & 2.178 & 14.851 & 0.774 & S0 & E & S0 & N & N \\
16 & 1237667324334637347 & 0.232 & 2.068 & 15.363 & 0.766 & SB0 & S0 & SBa & N & N \\
17 & 1237667324334702605 & 0.311 & 2.024 & 14.826 & 0.695 & SB0 & SB0/a & S0 & Y & Y \\
18 & 1237667324334702869 & 0.317 & 2.076 & 14.566 & 0.790 & SB0 & S0 & S0 & Y & Y \\
19 & 1237667324334833870 & 1.044 & 1.811 & 14.978 & 0.718 & S0 & S0 & - & N & N \\
20 & 1237667324334899374 & 1.195 & 1.751 & 14.745 & 0.734 & S0 & S0 & - & N & N \\
21 & 1237667443511525379 & 1.472 & 2.207 & 14.596 & 0.835 & S0 & E & - & N & N \\
22 & 1237667443511525432 & 1.360 & 2.302 & 14.191 & 0.833 & SB0 & E & - & N & N \\
23 & 1237667443511590950 & 1.202 & 2.319 & 14.008 & 0.836 & S0 & S0 & - & N & N \\
24 & 1237667443511590951 & 1.211 & 2.208 & 15.157 & 0.814 & SB0 & S0 & - & Y & Y \\
25 & 1237667443511722010 & 1.027 & 2.055 & 15.151 & 0.781 & SB0 & S0 & - & Y & Y \\
26 & 1237667443511787541 & 1.014 & 2.274 & 14.379 & 0.801 & SB0 & S0 & - & N & N \\
27 & 1237667444048396291 & 1.214 & 2.275 & 13.904 & 0.838 & SB0 & S0 & - & N & N \\
28 & 1237667444048461861 & 0.807 & 2.158 & 14.405 & 0.764 & S0 & S0 & - & N & N \\
29 & 1237667444048527399 & 0.577 & 2.270 & 14.206 & 0.796 & SB0 & E/S0 & S0 & Y & Y \\
30 & 1237667444048592990 & 0.298 & 2.177 & 14.285 & 0.815 & SB0 & SB0 & Sa & Y & Y \\
31 & 1237667444048593084 & 0.344 & 2.256 & 14.509 & 0.819 & SB0 & S0 & SB0 & Y & Y \\
32 & 1237667444048658449 & 0.225 & 2.076 & 15.302 & 0.776 & SB0 & S0 & SBa & N & N \\
33 & 1237667444048658521 & 0.120 & 2.302 & 14.486 & 0.817 & SB0 & E/S0 & SB0p & N & N \\
34 & 1237667444048658522 & 0.142 & 2.243 & 14.018 & 0.820 & SB0 & S0 & S0 & Y & Y \\
35 & 1237667444048658523 & 0.125 & 2.230 & 14.214 & 0.787 & S0 & S0 & E3 & N & N \\
36 & 1237667444048658535 & 0.093 & 1.907 & 14.900 & 0.798 & S0 & S0 & S0 & N & N \\
37 & 1237667444048658858 & 0.073 & 2.000 & 15.287 & 0.768 & SB0 & SB0 & SBa & N & Y \\
38 & 1237667444048724118 & 0.232 & 2.119 & 15.116 & 0.760 & SB0 & S0 & SB0 & N & Y \\
39 & 1237667444048724176 & 0.326 & 2.058 & 14.794 & 0.755 & SB0 & SB0 & SBa & Y & Y \\
40 & 1237667444048789721 & 0.581 & 1.964 & 14.917 & 0.753 & SB0 & S0 & S0 & Y & Y \\
41 & 1237667444048789764 & 0.615 & 2.272 & 13.819 & 0.801 & S0 & S0 & S0 & N & N \\
42 & 1237667444585201702 & 1.367 & 2.089 & 14.937 & 0.777 & SB0 & SB0 & - & Y & Y \\
43 & 1237667444585595001 & 0.704 & 2.204 & 13.687 & 0.791 & SB0 & S0 & S0 & N & N \\
44 & 1237667444585595059 & 0.724 & 2.097 & 14.614 & 0.761 & SB0 & S0 & S0 & N & N \\
45 & 1237667444585595093 & 0.792 & 2.238 & 13.705 & 0.803 & SB0 & S0 & SB0 & N & N \\
46 & 1237665440979026019 & 1.503 & 2.151 & 14.124 & 0.747 & S0 & - & - & N & N \\
47 & 1237665440979484734 & 2.393 & 1.962 & 15.312 & 0.751 & SB0 & - & - & N & N \\
48 & 1237665441516028062 & 2.486 & 1.907 & 15.574 & 0.760 & SB0 & - & - & N & Y \\
49 & 1237667253482553389 & 2.204 & 2.030 & 14.840 & 0.765 & SB0 & - & - & Y & Y \\
50 & 1237667322723827758 & 1.960 & 2.097 & 14.621 & 0.774 & S0 & - & - & N & N \\
51 & 1237667323260633139 & 1.507 & 2.026 & 15.253 & 0.776 & SB0 & - & - & N & Y \\
52 & 1237667324334964901 & 1.369 & 1.849 & 14.989 & 0.712 & SB0 & - & - & Y & Y \\
53 & 1237667324335030394 & 1.640 & 1.728 & 15.004 & 0.722 & S0 & - & - & N & N \\
54 & 1237667442974654524 & 1.831 & 2.405 & 12.984 & 0.834 & S0 & - & - & N & N \\
55 & 1237667442974720162 & 1.848 & 2.047 & 14.855 & 0.821 & SB0 & - & - & Y & Y \\
56 & 1237667443511591025 & 1.190 & 2.229 & 14.710 & 0.787 & S0 & - & - & N & N \\
57 & 1237667444048265289 & 1.560 & 2.071 & 15.056 & 0.797 & SB0 & - & - & Y & Y \\
58 & 1237667444048265310 & 1.751 & 1.649 & 15.092 & 0.741 & S0 & - & - & N & N \\
59 & 1237667444048330789 & 1.439 & 2.084 & 14.506 & 0.811 & SB0 & - & - & Y & Y \\
60 & 1237667444048658525 & 0.034 & 2.115 & 14.742 & 0.812 & SB0 & - & SB0 & N & Y \\
\hline
\end{tabular}
\end{table*}
\begin{table*}
\begin{minipage}[l]{15.57cm}
\contcaption{}
\end{minipage}
\begin{tabular}{lcrcccccccc}
\hline \hline
\multicolumn{1}{l}{ID} &  
\multicolumn{1}{c}{SDSS ID} & 
\multicolumn{1}{c}{$R_{\mathrm{proj}}$}  &
\multicolumn{1}{c}{$\log\sigma$}  &
\multicolumn{1}{c}{$r_{\mathrm{petro}}$} & 
\multicolumn{1}{c}{$g$-$r$} &
\multicolumn{1}{c}{Type} & 
\multicolumn{1}{c}{Type} &  
\multicolumn{1}{c}{Type} & 
\multicolumn{1}{c}{`ellipse'?} & 
\multicolumn{1}{c}{`ellipse'?} \\

\multicolumn{1}{l}{$\#$} &  
\multicolumn{1}{c}{(DR8)} & 
\multicolumn{1}{c}{Mpc}  &
\multicolumn{1}{c}{${\rm{km}\,\rm{s}^{-1}}$}  &
\multicolumn{1}{c}{mag} & 
\multicolumn{1}{c}{mag} &
\multicolumn{1}{c}{(This Study)} & 
\multicolumn{1}{c}{(D80)} &  
\multicolumn{1}{c}{(M08)} & 
\multicolumn{1}{c}{(strict)} & 
\multicolumn{1}{c}{(relaxed)} \\

\multicolumn{1}{l}{(1)} &  
\multicolumn{1}{c}{(2)} & 
\multicolumn{1}{c}{(3)}  &
\multicolumn{1}{c}{(4)}  &
\multicolumn{1}{c}{(5)} & 
\multicolumn{1}{c}{(6)} &
\multicolumn{1}{c}{(7)} & 
\multicolumn{1}{c}{(8)} &  
\multicolumn{1}{c}{(9)} & 
\multicolumn{1}{c}{(10)} & 
\multicolumn{1}{c}{(11)} \\

\hline
61 & 1237667444048658635 & 0.139 & 2.260 & 14.667 & 0.825 & SB0 & - & S0 & Y & Y \\
62 & 1237667444585005256 & 2.159 & 2.144 & 13.998 & 0.803 & SB0 & - & - & Y & Y \\
63 & 1237667444585922611 & 1.600 & 1.867 & 15.310 & 0.737 & SB0 & - & - & Y & Y \\
64 & 1237667444585922740 & 1.722 & 2.222 & 14.313 & 0.818 & SB0 & - & - & N & N \\
65 & 1237665024370999330 & 11.276 & 2.371 & 13.402 & 0.842 & SB0 & - & - & N & N \\
66 & 1237665024908722240 & 11.158 & 2.007 & 15.091 & 0.801 & SB0 & - & - & Y & Y \\
67 & 1237665225698377768 & 5.813 & 2.271 & 14.455 & 0.828 & S0 & - & - & N & N \\
68 & 1237665226774282293 & 10.053 & 2.199 & 13.821 & 0.751 & SB0 & - & - & N & Y \\
69 & 1237665428092944482 & 12.464 & 1.937 & 15.316 & 0.767 & S0 & - & - & N & N \\
70 & 1237665429164589088 & 5.150 & 2.370 & 12.895 & 0.808 & S0 & - & - & N & N \\
71 & 1237665443126116356 & 4.932 & 2.185 & 14.219 & 0.788 & S0 & - & - & N & N \\
72 & 1237665443126116437 & 4.906 & 2.311 & 13.615 & 0.854 & SB0 & - & - & N & N \\
73 & 1237665531707785219 & 13.908 & 2.422 & 13.598 & 0.840 & S0 & - & - & N & N \\
74 & 1237667255092838497 & 4.779 & 2.221 & 14.607 & 0.830 & SB0 & - & - & N & N \\
75 & 1237667321647661108 & 10.718 & 2.091 & 13.962 & 0.998 & SB0 & - & - & N & N \\
76 & 1237667322183942213 & 12.927 & 2.086 & 14.747 & 0.799 & SB0 & - & - & Y & Y \\
77 & 1237667322721730677 & 9.310 & 1.905 & 14.743 & 0.767 & SB0 & - & - & Y & Y \\
78 & 1237667322722320564 & 6.764 & 2.027 & 14.340 & 0.771 & SB0 & - & - & Y & Y \\
79 & 1237667442435752104 & 9.411 & 2.215 & 13.754 & 0.818 & SB0 & - & - & Y & Y \\
80 & 1237667442435817550 & 9.175 & 2.285 & 13.477 & 0.857 & S0 & - & - & N & N \\
81 & 1237667442436538434 & 6.463 & 2.130 & 14.247 & 0.799 & SB0 & - & - & N & N \\
82 & 1237667442437193733 & 4.263 & 2.147 & 14.527 & 0.843 & S0 & - & - & N & N \\
83 & 1237667443508576272 & 12.905 & 2.295 & 14.194 & 0.806 & S0 & - & - & N & N \\
\hline
\end{tabular}
\begin{minipage}[l]{15.57cm}
\small
(1) Galaxy ID for this study.
(2) SDSS DR8 object ID.
(3) Projected cluster radius.
(4) Central velocity dispersion.
(5) SDSS $r$-band magnitude using the AB system.
(6) SDSS $g$-$r$ colour.
(7) Hubble type as determined using the bar detection criteria in Section \ref{bar_criteria}.
(8) Hubble type as determined by \citet{Dressler80}.
(9) Hubble type as determined by \citet{Michard08}.
(10) Yes/No to whether a bar was detected using the ellipse fitting of galaxy isophotes, using strict detection criteria.
(11) Yes/No to whether a bar was detected using the ellipse fitting of galaxy isophotes, using relaxed detection criteria.
\end{minipage}
\end{table*}

\clearpage
\begin{table*}
\begin{minipage}[l]{17.44cm}
\caption{Bulge+Disk+Bar decomposition parameters for the $83$ S0s in this investigation. The parameters are for final accepted fitting stages only, i.e. bulge+disk+bar for barred lenticulars (SB0), and bulge+disk for unbarred lenticulars (S0). $\#1$-$64$ are the central sample, and $\#65$-$83$ are the outskirts sample. \label{params_table}}
\end{minipage}
\begin{tabular}{lcccccccccccccc}
\hline \hline

\multicolumn{1}{l}{ID} &
\multicolumn{1}{c}{ $\Delta \mathrm{RFF}$ } &
\multicolumn{1}{c}{ $B/T$ } &
\multicolumn{1}{c}{ $Bar/T$ } &
\multicolumn{1}{c}{ $r_{B}$ } &
\multicolumn{1}{c}{ $r_{D}$ } &
\multicolumn{1}{c}{ $r_{Bar}$ } &
\multicolumn{1}{c}{ $n_{B}$ } &
\multicolumn{1}{c}{ $n_{Bar}$ } &
\multicolumn{1}{c}{ $(b/a)_{B}$ } &
\multicolumn{1}{c}{ $(b/a)_{D}$ } &
\multicolumn{1}{c}{ $(b/a)_{Bar}$ } &
\multicolumn{1}{c}{ PA$_{B}$ } &
\multicolumn{1}{c}{ PA$_{D}$ } &
\multicolumn{1}{c}{ PA$_{Bar}$ } \\

\multicolumn{1}{l}{     $\#$      } &
\multicolumn{1}{c}{ \% } &
\multicolumn{1}{c}{ \% } &
\multicolumn{1}{c}{  \%  } &
\multicolumn{1}{c}{ kpc } &
\multicolumn{1}{c}{ kpc } &
\multicolumn{1}{c}{ kpc } &
\multicolumn{1}{c}{} &
\multicolumn{1}{c}{} &
\multicolumn{1}{c}{} &
\multicolumn{1}{c}{} &
\multicolumn{1}{c}{} &
\multicolumn{1}{c}{ $^{\circ}$ } &
\multicolumn{1}{c}{ $^{\circ}$ } &
\multicolumn{1}{c}{ $^{\circ}$ } \\

 \multicolumn{1}{l}{(1)} &\multicolumn{1}{c}{(2)} &\multicolumn{1}{c}{(3)} &\multicolumn{1}{c}{(4)} &
 \multicolumn{1}{c}{(5)} &\multicolumn{1}{c}{(6)} &\multicolumn{1}{c}{(7)} &\multicolumn{1}{c}{(8)} &
 \multicolumn{1}{c}{(9)} &\multicolumn{1}{c}{(10)} &\multicolumn{1}{c}{(11)} &\multicolumn{1}{c}{(12)} 
  &\multicolumn{1}{c}{(13)} &\multicolumn{1}{c}{(14)} &\multicolumn{1}{c}{(15)}  \\

\hline
01 & 0.22 & 42.56 & - & 0.58 & 3.93 & - & 1.77 & - & 0.51 & 0.50 & - & 40.55 & 41.60 & - \\
02 & 0.1 & 22.51 & - & 0.89 & 5.03 & - & 1.94 & - & 0.93 & 0.94 & - & 45.92 & 36.05 & - \\
03 & 1.49 & 24.76 & 13.00 & 0.70 & 4.99 & 2.93 & 1.31 & 0.39 & 0.77 & 0.74 & 0.34 & -32.12 & -18.40 & -45.26 \\
04 & 0.79 & 42.35 & - & 1.79 & 8.34 & - & 4.06 & - & 0.80 & 0.64 & - & -12.16 & -57.00 & - \\
05 & 3.29 & 46.58 & 22.01 & 0.51 & 3.23 & 2.38 & 2.45 & 0.45 & 0.65 & 0.74 & 0.34 & -7.66 & 9.42 & 4.84 \\
06 & 1.63 & 14.57 & 16.44 & 0.53 & 3.33 & 2.48 & 0.97 & 0.27 & 0.51 & 0.82 & 0.28 & -65.64 & -54.35 & -68.75 \\
07 & 1.81 & 9.59 & 5.67 & 0.31 & 3.50 & 0.96 & 0.97 & 0.35 & 0.61 & 0.54 & 0.57 & 19.33 & 38.12 & -6.19 \\
08 & 2.33 & 20.82 & 24.15 & 0.32 & 3.19 & 1.22 & 1.21 & 0.55 & 0.81 & 0.82 & 0.49 & -76.63 & -87.78 & -39.71 \\
09 & 2.99 & 31.91 & 14.38 & 0.86 & 4.44 & 2.91 & 1.81 & 0.25 & 0.74 & 0.78 & 0.39 & -12.33 & -28.57 & 19.82 \\
10 & 1.37 & 20.57 & 33.17 & 0.54 & 3.05 & 2.38 & 1.42 & 0.56 & 0.74 & 0.76 & 0.39 & 31.61 & 35.68 & 17.42 \\
11 & 2.52 & 32.50 & 33.14 & 0.89 & 6.83 & 3.11 & 1.99 & 0.78 & 0.83 & 0.71 & 0.63 & -37.77 & -24.98 & 16.89 \\
12 & 1.9 & 22.89 & 18.67 & 0.30 & 2.49 & 1.47 & 1.32 & 0.46 & 0.76 & 0.97 & 0.59 & -77.73 & 81.70 & 39.94 \\
13 & 1.89 & 46.88 & 11.19 & 0.57 & 4.11 & 3.06 & 1.69 & 0.40 & 0.74 & 0.63 & 0.27 & 50.32 & 46.55 & 45.92 \\
14 & 1.42 & 13.69 & 16.38 & 0.32 & 4.13 & 1.38 & 0.92 & 0.54 & 0.78 & 0.75 & 0.52 & -12.99 & -23.82 & 40.49 \\
15 & 1.02 & 69.29 & - & 1.54 & 3.03 & - & 2.95 & - & 0.59 & 0.63 & - & -78.56 & -83.62 & - \\
16 & 1.54 & 14.79 & 30.56 & 0.18 & 3.90 & 0.81 & 1.71 & 0.68 & 0.64 & 0.54 & 0.69 & -31.74 & -20.56 & 20.21 \\
17 & 1.56 & 17.36 & 20.14 & 0.48 & 4.77 & 3.28 & 1.58 & 0.64 & 0.64 & 0.64 & 0.34 & -13.03 & -0.01 & -19.80 \\
18 & 0.65 & 26.50 & 19.04 & 0.78 & 5.47 & 2.97 & 1.68 & 0.71 & 0.61 & 0.67 & 0.37 & -51.21 & -52.35 & -46.22 \\
19 & 0.11 & 11.77 & - & 0.46 & 5.13 & - & 1.22 & - & 0.57 & 0.46 & - & 53.77 & -88.82 & - \\
20 & 0.07 & 22.27 & - & 1.10 & 4.68 & - & 1.85 & - & 0.51 & 0.47 & - & 38.46 & 38.21 & - \\
21 & 0.79 & 34.71 & - & 0.44 & 3.26 & - & 1.56 & - & 0.78 & 0.85 & - & -76.95 & -44.47 & - \\
22 & 1.34 & 35.05 & 15.01 & 0.54 & 3.84 & 1.63 & 2.11 & 0.62 & 0.88 & 0.95 & 0.55 & -79.98 & -25.72 & 53.15 \\
23 & 0.57 & 74.54 & - & 1.46 & 3.57 & - & 2.60 & - & 0.88 & 0.97 & - & -14.90 & -16.64 & - \\
24 & 3.6 & 50.50 & 24.64 & 0.62 & 3.78 & 2.31 & 2.82 & 0.35 & 0.67 & 0.66 & 0.41 & -45.92 & -37.00 & -49.66 \\
25 & 1.35 & 29.94 & 23.67 & 0.47 & 4.02 & 2.40 & 1.46 & 0.55 & 0.56 & 0.66 & 0.37 & -20.22 & -10.01 & -23.20 \\
26 & 1.59 & 36.32 & 7.73 & 0.42 & 5.08 & 1.87 & 1.68 & 0.34 & 0.82 & 0.73 & 0.39 & 41.15 & 42.21 & 37.83 \\
27 & 0.79 & 28.23 & 9.60 & 0.55 & 7.74 & 1.85 & 2.39 & 0.36 & 0.93 & 0.97 & 0.77 & 30.78 & -18.15 & 78.01 \\
28 & 0.89 & 45.23 & - & 0.91 & 4.31 & - & 3.35 & - & 0.76 & 0.65 & - & -31.41 & -18.38 & - \\
29 & 1.63 & 53.73 & 14.73 & 1.30 & 6.69 & 2.09 & 3.87 & 0.37 & 0.92 & 0.89 & 0.58 & 58.93 & 81.70 & 42.72 \\
30 & 4.1 & 35.95 & 17.37 & 0.79 & 5.27 & 4.33 & 2.26 & 0.41 & 0.94 & 0.75 & 0.34 & -56.23 & -54.33 & -70.48 \\
31 & 2.71 & 20.20 & 9.67 & 0.31 & 2.80 & 1.12 & 0.85 & 0.73 & 0.95 & 0.95 & 0.42 & 70.18 & 37.79 & -20.83 \\
32 & 2.23 & 15.71 & 18.48 & 0.15 & 1.90 & 0.63 & 0.76 & 0.43 & 0.83 & 0.70 & 0.64 & -65.51 & -66.41 & 11.69 \\
33 & 2.17 & 29.97 & 22.32 & 0.38 & 3.81 & 1.37 & 1.60 & 0.46 & 0.83 & 0.61 & 0.77 & 16.77 & 20.18 & -53.58 \\
34 & 3.59 & 23.17 & 23.40 & 0.53 & 7.06 & 3.42 & 1.77 & 0.61 & 0.85 & 0.80 & 0.43 & 65.53 & 62.24 & 85.30 \\
35 & 0.14 & 25.04 & - & 0.53 & 3.58 & - & 2.32 & - & 0.86 & 0.71 & - & 35.75 & 5.43 & - \\
36 & 0.19 & 24.37 & - & 0.85 & 4.15 & - & 2.09 & - & 0.64 & 0.63 & - & -52.02 & -46.78 & - \\
37 & 3.34 & 13.31 & 22.53 & 0.41 & 4.65 & 2.58 & 1.47 & 0.36 & 0.69 & 0.54 & 0.39 & -31.11 & -42.29 & -2.79 \\
38 & 1.46 & 30.99 & 7.86 & 0.33 & 3.70 & 1.52 & 1.27 & 0.32 & 0.81 & 0.63 & 0.44 & -72.53 & -82.81 & -31.81 \\
39 & 1.9 & 20.47 & 12.07 & 0.37 & 3.85 & 1.58 & 1.48 & 0.48 & 0.78 & 0.80 & 0.34 & -16.67 & 1.84 & -47.72 \\
40 & 1.38 & 17.55 & 21.31 & 0.43 & 4.52 & 1.71 & 1.45 & 0.86 & 0.85 & 0.90 & 0.41 & 24.51 & 22.75 & 32.93 \\
41 & 1.03 & 33.58 & - & 0.73 & 4.32 & - & 1.73 & - & 0.71 & 0.55 & - & 45.78 & 47.50 & - \\
42 & 2.92 & 29.10 & 11.25 & 0.36 & 3.04 & 1.90 & 1.45 & 0.27 & 0.85 & 0.81 & 0.36 & 49.02 & 33.45 & 4.68 \\
43 & 0.73 & 16.34 & 6.55 & 0.52 & 5.59 & 2.15 & 1.73 & 0.44 & 0.65 & 0.56 & 0.40 & 89.14 & -87.94 & 84.21 \\
44 & 1.84 & 24.05 & 25.60 & 0.44 & 3.64 & 1.66 & 1.71 & 0.42 & 0.53 & 0.54 & 0.39 & -41.70 & -46.01 & -46.66 \\
45 & 1.64 & 12.64 & 17.31 & 0.32 & 4.91 & 1.09 & 1.19 & 0.54 & 0.74 & 0.55 & 0.67 & -72.38 & -71.21 & 38.56 \\
46 & 0.14 & 70.15 & - & 30.15 & 5.68 & - & 17.55 & - & 0.61 & 0.44 & - & 88.79 & 86.68 & - \\
47 & 0.93 & 12.28 & 21.33 & 0.22 & 3.12 & 0.68 & 1.02 & 0.77 & 0.40 & 0.47 & 0.73 & -80.31 & 86.41 & 51.71 \\
48 & 1.58 & 17.25 & 17.62 & 0.24 & 2.74 & 1.07 & 0.50 & 0.35 & 0.72 & 0.72 & 0.42 & 36.48 & 14.28 & 84.28 \\
49 & 1.28 & 21.68 & 11.54 & 0.26 & 4.02 & 1.16 & 1.76 & 0.33 & 0.91 & 0.86 & 0.55 & -79.53 & -67.67 & 62.87 \\
50 & 0.82 & 39.39 & - & 0.52 & 4.76 & - & 2.38 & - & 0.73 & 0.58 & - & 33.92 & 35.42 & - \\
51 & 2.25 & 10.50 & 15.15 & 0.29 & 3.85 & 1.46 & 0.76 & 0.35 & 0.67 & 0.45 & 0.37 & -0.68 & 9.98 & -23.83 \\
52 & 2.54 & 7.30 & 9.61 & 0.28 & 4.23 & 2.72 & 0.79 & 0.20 & 0.74 & 0.58 & 0.22 & -25.33 & -55.89 & -30.31 \\
53 & 0.43 & 56.05 & - & 5.40 & 3.33 & - & 6.22 & - & 0.61 & 0.95 & - & -41.68 & -62.03 & - \\
54 & 0.32 & 98.59 & - & 14.96 & 10.84 & - & 7.25 & - & 0.76 & 0.23 & - & -46.38 & 64.12 & - \\
55 & 3.43 & 34.66 & 11.84 & 0.58 & 5.35 & 3.27 & 2.01 & 0.31 & 0.93 & 0.71 & 0.33 & -49.20 & 18.03 & 28.32 \\
56 & 0.52 & 74.24 & - & 1.19 & 2.61 & - & 4.60 & - & 0.60 & 0.50 & - & -42.14 & -33.32 & - \\
57 & 1.54 & 25.22 & 13.79 & 0.25 & 3.62 & 1.23 & 1.47 & 0.47 & 0.91 & 0.80 & 0.46 & 71.98 & 72.04 & 48.69 \\
58 & 0.06 & 6.09 & - & 0.53 & 3.71 & - & 2.56 & - & 0.81 & 0.75 & - & 27.66 & 48.77 & - \\
59 & 3.59 & 19.99 & 14.09 & 0.50 & 6.46 & 2.51 & 1.36 & 0.49 & 0.72 & 0.73 & 0.33 & -54.72 & -46.44 & -21.17 \\
\hline
\end{tabular}
\end{table*}
\begin{table*}
\begin{minipage}[l]{17.22cm}
\contcaption{}
\end{minipage}
\begin{tabular}{lcccccccccccccc}
\hline \hline

\multicolumn{1}{l}{ID} &
\multicolumn{1}{c}{ $\Delta \mathrm{RFF}$ } &
\multicolumn{1}{c}{ $B/T$ } &
\multicolumn{1}{c}{ $Bar/T$ } &
\multicolumn{1}{c}{ $r_{B}$ } &
\multicolumn{1}{c}{ $r_{D}$ } &
\multicolumn{1}{c}{ $r_{Bar}$ } &
\multicolumn{1}{c}{ $n_{B}$ } &
\multicolumn{1}{c}{ $n_{Bar}$ } &
\multicolumn{1}{c}{ $(b/a)_{B}$ } &
\multicolumn{1}{c}{ $(b/a)_{D}$ } &
\multicolumn{1}{c}{ $(b/a)_{Bar}$ } &
\multicolumn{1}{c}{ PA$_{B}$ } &
\multicolumn{1}{c}{ PA$_{D}$ } &
\multicolumn{1}{c}{ PA$_{Bar}$ } \\

\multicolumn{1}{l}{     $\#$      } &
\multicolumn{1}{c}{ \% } &
\multicolumn{1}{c}{ \% } &
\multicolumn{1}{c}{  \%  } &
\multicolumn{1}{c}{ kpc } &
\multicolumn{1}{c}{ kpc } &
\multicolumn{1}{c}{ kpc } &
\multicolumn{1}{c}{} &
\multicolumn{1}{c}{} &
\multicolumn{1}{c}{} &
\multicolumn{1}{c}{} &
\multicolumn{1}{c}{} &
\multicolumn{1}{c}{ $^{\circ}$ } &
\multicolumn{1}{c}{ $^{\circ}$ } &
\multicolumn{1}{c}{ $^{\circ}$ } \\

 \multicolumn{1}{l}{(1)} &\multicolumn{1}{c}{(2)} &\multicolumn{1}{c}{(3)} &\multicolumn{1}{c}{(4)} &
 \multicolumn{1}{c}{(5)} &\multicolumn{1}{c}{(6)} &\multicolumn{1}{c}{(7)} &\multicolumn{1}{c}{(8)} &
 \multicolumn{1}{c}{(9)} &\multicolumn{1}{c}{(10)} &\multicolumn{1}{c}{(11)} &\multicolumn{1}{c}{(12)} 
  &\multicolumn{1}{c}{(13)} &\multicolumn{1}{c}{(14)} &\multicolumn{1}{c}{(15)}  \\

\hline
60 & 2.46 & 12.94 & 11.71 & 0.28 & 3.82 & 1.44 & 1.06 & 0.52 & 0.79 & 0.68 & 0.39 & -73.10 & -82.97 & -27.46 \\
61 & 1.16 & 56.90 & 10.22 & 0.80 & 3.75 & 2.26 & 2.67 & 0.39 & 0.84 & 0.78 & 0.44 & 12.00 & 15.92 & 26.53 \\
62 & 4.25 & 14.20 & 14.63 & 0.57 & 8.08 & 3.35 & 1.61 & 0.30 & 0.73 & 0.68 & 0.34 & -33.41 & -26.83 & -78.54 \\
63 & 0.53 & 15.04 & 13.00 & 0.18 & 2.57 & 0.79 & 1.13 & 0.80 & 0.66 & 0.85 & 0.34 & 48.76 & -0.80 & 54.90 \\
64 & 1.13 & 27.31 & 20.53 & 0.52 & 6.32 & 2.17 & 2.33 & 0.56 & 0.85 & 0.82 & 0.76 & -31.92 & -45.74 & 87.96 \\
65 & 0.67 & 31.74 & 13.68 & 1.00 & 8.72 & 4.26 & 2.25 & 0.36 & 0.75 & 0.71 & 0.54 & -44.39 & -45.02 & -41.57 \\
66 & 1.87 & 14.41 & 14.15 & 0.30 & 2.96 & 1.48 & 0.82 & 0.64 & 0.86 & 0.93 & 0.31 & -24.27 & 76.54 & -9.38 \\
67 & 0.37 & 38.63 & - & 0.48 & 4.28 & - & 1.93 & - & 0.91 & 0.79 & - & 23.93 & -53.27 & - \\
68 & 1.54 & 28.18 & 15.73 & 0.38 & 5.78 & 1.80 & 1.43 & 0.32 & 0.63 & 0.61 & 0.55 & -76.94 & -76.90 & -66.97 \\
69 & 0.5 & 33.46 & - & 0.42 & 3.09 & - & 4.03 & - & 0.64 & 0.71 & - & -79.84 & 47.28 & - \\
70 & 0.18 & 60.90 & - & 2.65 & 10.04 & - & 3.43 & - & 0.76 & 0.63 & - & 71.86 & 74.38 & - \\
71 & 0.89 & 51.88 & - & 0.86 & 5.19 & - & 1.69 & - & 0.68 & 0.76 & - & 71.66 & 79.23 & - \\
72 & 0.73 & 23.48 & 5.66 & 0.42 & 4.28 & 1.68 & 1.39 & 0.40 & 0.88 & 0.89 & 0.57 & 63.69 & 73.73 & 85.61 \\
73 & 0.68 & 28.70 & - & 0.55 & 4.13 & - & 2.55 & - & 0.90 & 0.86 & - & 87.44 & 84.22 & - \\
74 & 1.3 & 20.16 & 8.20 & 0.30 & 3.00 & 1.07 & 0.97 & 0.45 & 0.83 & 0.74 & 0.48 & -37.07 & -30.43 & -77.14 \\
75 & 1.31 & 19.47 & 13.36 & 0.44 & 6.08 & 1.86 & 2.43 & 0.35 & 0.90 & 0.96 & 0.78 & 4.06 & -7.54 & 41.40 \\
76 & 1.93 & 17.16 & 22.64 & 0.29 & 5.31 & 1.64 & 1.34 & 0.53 & 0.88 & 0.87 & 0.65 & -37.71 & -34.29 & 48.26 \\
77 & 3.14 & 15.54 & 21.26 & 0.60 & 5.49 & 2.82 & 1.24 & 0.50 & 0.86 & 0.84 & 0.31 & -29.67 & -51.36 & -19.93 \\
78 & 1.15 & 5.59 & 7.46 & 0.37 & 4.97 & 2.36 & 0.41 & 0.30 & 0.53 & 0.75 & 0.14 & 26.06 & 22.69 & 29.64 \\
79 & 2.2 & 15.03 & 12.11 & 0.36 & 5.54 & 2.20 & 1.42 & 0.72 & 0.95 & 0.93 & 0.40 & 9.06 & 22.58 & 5.68 \\
80 & 0.03 & 71.03 & - & 3.66 & 19.77 & - & 4.12 & - & 0.93 & 0.61 & - & 19.29 & 27.95 & - \\
81 & 3.56 & 25.56 & 32.04 & 0.55 & 6.54 & 3.31 & 1.51 & 0.83 & 0.82 & 0.96 & 0.67 & 12.55 & -44.18 & -62.94 \\
82 & 1.43 & 67.89 & - & 1.81 & 8.41 & - & 4.40 & - & 0.78 & 0.61 & - & -6.86 & -19.52 & - \\
83 & 0.47 & 45.86 & - & 0.80 & 4.23 & - & 1.75 & - & 0.62 & 0.74 & - & -52.89 & -53.22 & - \\
\hline
\end{tabular}
\begin{minipage}[l]{17.22cm}
\small
(1) Galaxy ID for this study.
(2) The change in residual flux fraction when a bar component is added to the model (see Section \ref{bar_criteria}).
(3) Bulge light fraction.
(4) Bar light fraction.
(5) Bulge effective radius.
(6) Disk effective radius.
(7) Bar effective radius.
(8) Bulge S\'ersic index.
(9) Bar S\'ersic index.
(10) Bulge axial ratio.
(11) Disk axial ratio.
(12) Bar axial ratio.
(13) Bulge position angle.
(14) Disk position angle.
(15) Bar position angle.
\end{minipage}
\end{table*}


\end{document}